\documentclass[usenatbib]{mn2e}

\usepackage{graphicx,color,here}
\usepackage{natbib}
\usepackage{amssymb}
\usepackage{amsmath}

\author[R.~P. Eatough et al.]{R.~P. Eatough$^{1,2}$\thanks{E-mail:
reatough@mpifr-bonn.mpg.de}, M. Kramer$^{1,2}$, A.~G. Lyne$^2$ and
M. J. Keith$^{3}$
\\$^1$ Max-Planck-Institut f\"ur Radioastronomie, Auf dem H\"ugel 69,
53121, Bonn, Germany.  
\\$^2$ Jodrell Bank Centre for Astrophysics, Alan Turing Building, 
University of Manchester, Manchester, M13 9PL, United Kingdom.
\\$^3$ Australia Telescope National Facility,
CSIRO Astronomy \& Space Science, PO Box 76, Epping, NSW 1710, Australia.}
\date{1 May 2008}

\title{A coherent acceleration search of the Parkes multi-beam pulsar
  survey - techniques and the discovery and timing of 16 pulsars}

\begin{document}

\maketitle

\vspace{3cm}
\begin{abstract}
A fully coherent acceleration search algorithm has been applied to the
Parkes multi-beam pulsar survey of the Galactic plane to search for
previously undiscovered relativistic binary pulsars. The search has
resulted in the discovery of 16 pulsars including a binary millisecond
pulsar and an intermittent pulsar. Despite a number of promising
candidates there have been no new discoveries of relativistic binary
pulsars. Here we detail the acceleration search performed in our
analysis and present coherent timing solutions for each of pulsars
discovered. In light of the lack of discoveries of relativistic binary
pulsars, we also discuss the technique of acceleration searching and
its effectiveness in finding these systems.
\end{abstract}


\vspace{3cm}

\pagenumbering{arabic}
\pagestyle{plain}

\section{Introduction}
\label{s:intro}
Radio pulsars in compact, highly-relativistic binary systems, such as
double neutron star systems (DNS) and pulsar white dwarf systems
(NS-WD), provide the most precise tests of Einstein's theory of
General Relativity (GR) and alternative theories of gravity in the
strong-field regime (e.g. \citeauthor{tw89}~1989,
\citeauthor{ksm+06}~2006, \citeauthor{bkk+08}~2008,
\citeauthor{fwef+12}~2012). Future more stringent tests of gravity
could be performed with more compact and/or higher-mass systems such
as pulsar-black hole binary systems (NS-BH)
(\citeauthor{Kramer:2004}~2004), or even pulsars orbiting closely
around the super massive black hole at the centre of the Galaxy,
Sgr~A$^{\star}$ (\citeauthor{Pfahl:2005}~2005,
\citeauthor{lwk+12}~2012). The increased orbital velocities and deeper
gravitational potentials present in such systems would result in
larger post-Keplerian (PK) orbital effects, which are measurable
through pulsar timing (\citeauthor{de98}~1998,
\citeauthor{wk99}~1999).

Searches for these highly-prized systems are notoriously difficult.
After correction for the frequency dependent dispersion in the arrival
time of pulses, caused by the unknown free electron content along the
line-of-sight (`dedispersion'), standard pulsar searches look for
significant features in the fluctuation spectrum of the dedispersed time
series (see, for example, \citeauthor{lk05}~2005, for a detailed
  description of pulsar search methods).  In relativistic binary
systems, Doppler smearing of the pulse frequency, caused by rapid
orbital motion during the survey observation, renders standard
Fourier-based pulsar search algorithms ineffective (see, e.g.
\citeauthor{jk91}~1991). The binary search algorithms that compensate
for these effects are computationally expensive in blind pulsar
searches. Hence, the analysis of large-scale pulsar surveys, like the
Parkes multi-beam pulsar survey (PMPS), have not yet been performed
with optimum binary searches (\citeauthor{fsk+04}~2004).  In this work
we have attempted to further combat binary selection effects in the
Parkes multi-beam pulsar survey by re-processing the entire survey
with an efficient `coherent acceleration search'.

The outline of the rest of this paper is as follows: In Section 2, we
give a summary of the PMPS and describe previous binary searches of
the survey data. Section 3 is a description of the binary search
algorithm employed in this search. In Section 4, details of the
re-processing data pipeline are given. Section 5 presents timing
solutions of all 16 pulsars and further details on any of the notable
discoveries. Section 6 provides a discussion of possible reasons for
the lack of discoveries of any new relativistic binaries in addition
to further tests of the acceleration algorithm presented here. Section
7 is a summary and a discussion of future work on searches for
relativistic binary pulsars.

\section{The Parkes multi-beam pulsar survey}
\label{s:pmps}
The PMPS is the most successful pulsar survey ever completed, having
discovered 770 pulsars up to the start of this work
(\citeauthor{mlc+01}~2001, \citeauthor{mhl+02}~2002,
\citeauthor{kbm+03}~2003, \citeauthor{hfs+04}~2004,
\citeauthor{fsk+04}~2004, \citeauthor{lfl+06}~2006,
\citeauthor{kel+09}~2009) and over 30 pulsars discovered by
their single pulses (\citeauthor{mll+06}~2006,
\citeauthor{kle+10}~2010, \citeauthor{kkl+11}~2011). The survey was
carried out between 1997 and 2003 using a 13-beam receiver with a
bandwidth of 288 MHz centred on 1374 MHz and deployed on the 64-m
Parkes radio telescope in Australia.  It covers a strip along the
Galactic plane from $l = 260^{\circ}$ to $l = 50^{\circ}$ and $|b| <
5^{\circ}$ with time series of 35 minutes digitally sampled every
$250\;\mu{\rm s}$. Full details of the survey can be found in
\citeauthor{mlc+01}~(2001).

\subsection{Previous binary searches}
\label{s:previous_searches}
Despite the early successes of the PMPS the number of compact binary
pulsars (orbital period, $P_{\rm b}<1$ day) and millisecond pulsars
(spin period, $P<30$ ms) discovered was low; one and three respectively
(Table~3.2, \citeauthor{fau04}~2004). The long integration time of 35 minutes
gave the PMPS unprecedented sensitivity to long-period pulsars,
although, in searches for relativistic binary pulsars, where the spin
frequency can change over the duration of the observation, sensitivity
was lost in standard Fourier periodicity searches due to smearing of
the spectral features of the pulsar.

After an improvement in computational resources at the Jodrell Bank
Observatory, in the form of the COBRA 182-node Beowulf cluster, a full
re-processing of the PMPS was performed using two search algorithms
designed to address the previously-described binary selection effects
(\citeauthor{fau04}~2004, \citeauthor{fsk+04}~2004). The algorithms
used in this analysis were the `stack-slide search' (see, for example,
\citeauthor{wnh+91}~1991) and the `phase modulation search'
(\citeauthor{rce03}~2003). In the stack-slide search technique each
dedispersed time series is split into a number of contiguous segments
(16 in the case of the PMPS re-processing), which are Fourier
transformed separately.  The resulting fluctuation spectra are then
summed together with different linear offsets to correct for linear
drift in pulse frequency caused by any constant line-of-sight
acceleration. So called `acceleration searches' have been shown
  to be valid for observations where the integration time, $T_{\rm
    obs}$, is $\lesssim P_{\rm b}/10$ (\citeauthor{rce03}~2003). For
the PMPS this corresponded to binary systems with $P_{\rm b}\gtrsim6$
hr.

The phase modulation search works by detecting ``periodic sidebands''
in the fluctuation spectrum; the common marker of a pulsar in a
binary system (\citeauthor{rce03}~2003). Pulsars in binary systems
where at least $1.5$ orbits are completed during the observation time
can be detected with this technique. For the PMPS this corresponded
to systems with $P_{\rm b}\lesssim0.4$ hr.

The \citeauthor{fsk+04}~(2004) reprocessing found 11 binary pulsars
including the double neutron star system, PSR J1756$-$2251, that would
not have been discovered without the stack-slide search
(\citeauthor{fkl+05}~2005).  Of the other 10 binary systems, three
were detected with higher significance because of the stack-slide
search (on average by a factor of $\sim1.9$: Table 7.2,
\citeauthor{fau04}~2004). Results from the phase modulation search
have never been fully inspected due to difficulties in reconstructing
the predicted orbit of pulsar candidates (Faulkner, private
communication).

Although the stack-slide search method improved sensitivity to pulsars
in some fast binary systems, sensitivity was lost due to the
incoherent summation of spectra.  By splitting the time series into a
number of segments, each of which is Fourier transformed and then
summed, the phase coherence of any real signal is not preserved
throughout the entire length of the observation.  In the presence of
Gaussian noise, and assuming stacks of equal length, the
signal-to-noise ratio (S/N) of a real signal increases $\propto
\sqrt[4]n$ where $n$ is the number of spectra stacked
(\citeauthor{2000PhRvD..61h2001B}~2000). As already noted, the
stack-slide and phase modulation searches left a gap in sensitivity
for pulsars in binary systems with $P_{\rm b}$ between $\sim0.4$ and
$\sim6$ hr. For systems that lie in this region of parameter space,
sensitivity might also have been lost because of observed changes in
acceleration (`jerk') that the linear offsets applied in the
stack-slide search cannot compensate for.  Both the orbital-period
sensitivity gap and the reduction in sensitivity due to the incoherent
nature of the stack search might have been critical in the detection
of a relativistic binary pulsar that was near the detection threshold
of the survey.  In this work we describe further efforts to tackle
binary selection effects in the PMPS.

\section{A coherent acceleration search algorithm, {\sc PMACCN}}
\label{s:algorithm}
It is possible to perform acceleration searches in the time domain
that are fully coherent and without the loss of sensitivity
encountered in stacking algorithms. One such technique is `time
domain resampling'; a method that enabled the discovery of nine
binary millisecond pulsars in observations of 47 Tucanae at 20-cm
\citep{clf+00}. Following the description in Section 2 of
\citeauthor{clf+00}~(2000), the time series is transformed into a
frame which is inertial with respect to the pulsar by application of the
Doppler formula to relate a time interval in the pulsar frame $\tau$
to the time interval in the observed frame $t$,
\begin{equation}
\tau (t) = \tau_{0}\left(1+v(t)/c \right ),
\end{equation}
where $v(t)$ is the changing line-of-sight velocity of the pulsar, $c$
is the speed of light, and the constant $\tau_{0}$ is chosen such that
$\tau$ is equivalent to the sampling interval, $\tau_{\rm samp}$ at
the midpoint of the observation. New samples are computed from a
linear interpolation over the original time series
\citep{mk84}. Following resampling, the time series can be searched
with standard Fourier techniques, as if the pulsar were isolated (see
for example \citeauthor{lk05}~2005). To fully remove the effects of
orbital motion, the exact form of $v(t)$ must be known, although in a
blind search, where the orbital elements are a priori unknown, this
would require a search over five Keplerian parameters; a task that
would require significant computational resources (see for example,
\citeauthor{Dhurandhar:2001}~2001, and discussion in Appendix A). To
reduce the amount of computation required, the assumption of a
constant line-of-sight acceleration, $a$, can once again be applied,
so that $v(t)=at$. Now, searches in just one parameter, constant
line-of-sight acceleration, $a$, can be made, but in a fully coherent
manner.

\begin{figure}
\centering
\includegraphics[scale=0.65]{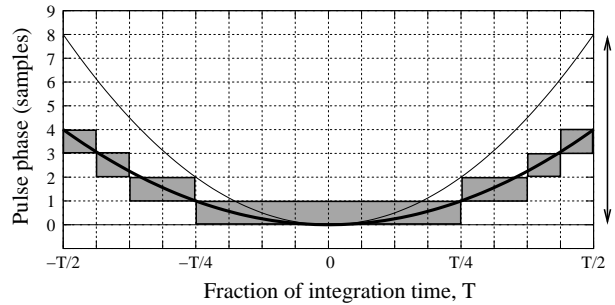}
\caption{Time versus pulse phase for an idealized top hat pulse
  with a width equal to the sampling interval (dark grey blocks) and a
  true acceleration, indicated by the thick black line, that lies
  directly in between acceleration trials given by
  Equation~\ref{e:accstep} (thin black lines). Arbitrary time
  intervals are marked by the vertical dotted lines. The size of the
  acceleration step is indicated by the vertical arrow on the
  right. For signals that have a true acceleration that lies in
  between acceleration trials, at least 50 per cent of the pulses are
  smeared by less than $\tau_{\rm samp}$.}
\label{f:sample_broad}
\end{figure}

A step size between each acceleration trial, $\delta a$ was determined
by fixing the amount of signal advance or retardation with respect to
either end of the integration\footnote{\footnotesize The effect of the
  constant, $\tau_{0}$ in Equation 1 is such that resampling is
  performed with respect to the midpoint of the observation.} to a
value that was deemed acceptable. This value, which we shall now term
`pulse broadening', $\tau_{\rm accn}$, was chosen to be four times the
sampling interval, $\tau_{\rm samp}$, i.e.  if the pulsar signal fell
between two acceleration trials this would be the maximum value of
pulse broadening at the end points of the data. Because of the
quadratic nature of the pulse broadening with time, due to a constant
acceleration, $a$ (see for example Equation 4 in
\citeauthor{jk91}~1991),
\begin{equation}
\tau_{\rm accn}(t) = at^2/2c\rm{,}
\end{equation}
at least 50 per cent of pulses in the integration would be smeared in
time by less than one time sample. Setting the pulse broadening time,
given by Equation 2, equal to $8\,\tau_{\rm samp}$ allows a maximum of
$4\,\tau_{\rm samp}$ pulse smearing for signals that lie exactly
in-between acceleration trials.  Remembering that resampling is
performed with respect to the midpoint of the data, $t=T_{\rm obs}/2$,
giving the acceleration step size, $\delta a$:
\begin{equation}
\delta a = 64c\tau_{\rm samp}/T_{\rm obs}^2.
\label{e:accstep} 
\end{equation}
Our choice of acceleration step size is illustrated in
Figure~\ref{f:sample_broad}. Here we plot the pulse phase as a
function of time for an idealized top hat pulse that has a width equal
to a single sample, and that is accelerated to a value that is
directly in between acceleration trials. The pulse phase broadening at
the end points of the integration is therefore $4\,\tau_{\rm samp}$;
the maximum value it can take. For a broadening of $4\,\tau_{\rm
  samp}$ at the end points it can be seen that 50 per cent of the
pulses are smeared in phase by less than $\tau_{\rm
  samp}$. Although the total broadening time is less
  than $\tau_{\rm samp}$ for half of the observation, the majority of
  power in a sample will be smeared into the next sample when, for
  example, $\tau_{\rm accn} > 0.5 \,\tau_{\rm samp}$, which occurs
  after $T_{\rm obs}/\sqrt{32}$ using this step size.

The acceleration step size given in Equation~\ref{e:accstep} scales
linearly with the sampling interval and with inverse square of the
integration time. Longer integrations and smaller sample intervals
therefore require finer acceleration steps.  Both of these parameters
could be tuned within the limits of the PMPS data to find an optimum
acceleration step size which covered new regions of parameter space
without excessive levels of computation. The final processing
parameters of our search algorithm, termed PMACCN, are given in
Table~\ref{t:procparams}. Data processing benchmarks and algorithm
performance tests described in
\citeauthor{Eatough:thesis}~(2009)\footnote{
www.jb.man.ac.uk/$\sim$reatough/reatough\_thesis.pdf}
favoured a search based on two independent half segments of the
original 35 minute integration, and a sampling interval increased by a
factor of four from the original $250\;\mu{\rm s}$ to 1 ms. Using this
choice of processing parameters a wide range of accelerations ($\pm
500 \,{\rm m\;s^{-2}}$) could be searched in a manageable time on the
computational hardware available (see Section 4). The acceleration
range chosen is nearly twice the maximum acceleration present in the
most relativistic binary pulsar known: the Double Pulsar system
($\sim260\,{\rm m\;s^{-2}}$: \citeauthor{bdp+03}~2003,
\citeauthor{ksm+06}~2006) and is a value that might be expected in
compact NS-BH systems (see Section 6). Analyzing independent halves
also provided better sensitivity to more compact binary systems
($P_{\rm b} \gtrsim 3$ hr) that would have been missed by acceleration
searches of the full length observation, because of the effects of
jerk (e.g. $P_{\rm b} \gtrsim 6$ hr in \citeauthor{fsk+04}~2004). The
sampling interval of 1 ms was not optimal for millisecond pulsars,
however it is expected that in most NS-BH systems the pulsar will be
the second-born object and will not have undergone recycling
(\citeauthor{Sipior:2004}~2004).

\begin{table}
\caption[PMPS reprocessing parameters for a single
  beam.]{\textsc{PMACCN} processing parameters. The step size in
  dispersion measure (DM) was determined following standard procedures
  outlined in \citeauthor{lk05}~(2005). The range in DM searched was
  defined by the expected values in the Galactic plane given by the
  \citeauthor{cl02}~(2002) free electron density model of the
  Galaxy. Each data segment corresponds to one independent half of the
  original 35 minute integration.}
\label{t:procparams}
\centering
\vspace{10pt}
\begin{tabular}{lll}
\hline\hline \multicolumn{1}{c}{Parameter}  &
\multicolumn{1}{c}{Value}\\\hline
Integration time & 2$\times$1050 s\\
Sample interval & 1 ms\\
Number of samples per data segment & $2^{20}$\\
Size of the data segments & 49 MB\\
Acceleration search range & $\pm 500\;{\rm m\;s^{-2}}$\\
Number of acceleration trials & 59 \\
DM search range & 0$-$1656 cm$^{-3}$pc \\
Number of DM trials & 152 \\
Processing time (per data segment) & $\sim 2.25$ hr\\
Total number of data segments & 2$\times$34710\\
\hline
\end{tabular}
\end{table}

\begin{figure}
\centering
\includegraphics[scale=0.8]{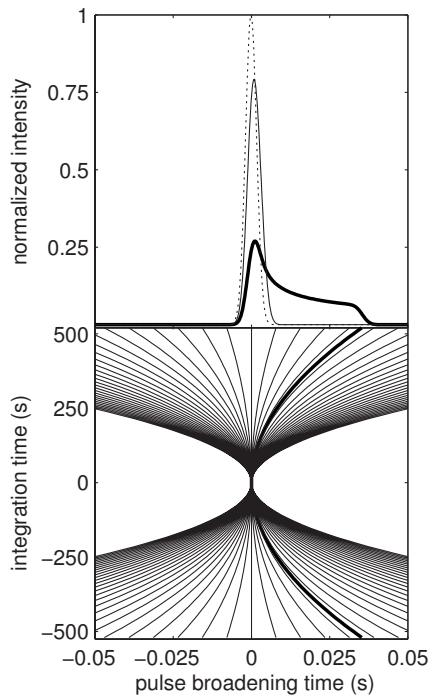}
\caption[Signal advance or retard due to a number of constant
  accelerations.]{The bottom panel shows the signal advance or retard
  due to a number of constant accelerations. The thin black lines
  represent accelerations for which the data can be corrected using
  the resampling method. The step size between acceleration trials is
  that given by Equation~3, for a sample interval of 1 ms and an
  integration time of 1050 s, as in the PMACCN search algorithm. In
  this case the simulated pulsar signal has an acceleration of
  $78.3\;{\rm m\;s^{-2}}$ and falls in the middle of two acceleration
  trials (thick black line). The upper panel shows the effect of the
  pulse broadening on a simulated Gaussian pulse for a pulsar with a
  spin period of 50 ms and a pulse width of 10 per cent. The dotted
  line shows the original pulse profile, with no broadening; the thick
  black line shows the profile broadening caused by the acceleration and
  the thin black line shows the ``acceleration-corrected'' profile for
  acceleration trials closest to the true acceleration.}
 \label{f:stepsize}
\end{figure}

Understanding the response of the data processing system to incorrect
acceleration trials is important for validating both genuinely
accelerated and even solitary pulsar candidates. The full effects of a
constant line-of-sight acceleration on the integrated pulse S/N can be
examined if we consider a simple Gaussian pulse convolved with the
pulse broadening function given by Equation 2.  In
Figure~\ref{f:stepsize}, the bottom panel shows the signal advance or
retard due to a number of constant accelerations within the 1050 s of
the PMACCN search algorithm. The thin black lines represent
accelerations that can be compensated for using acceleration steps
based on Equation~3.  Here we plot the ``worst case'' scenario where a
true signal from a pulsar, accelerated to $78.3\;{\rm m\;s^{-2}}$,
shown by the thick black line, lies exactly in the middle of two
acceleration trials. If resampling is performed with respect to the
midpoint of the data then the maximum pulse broadening is 4 ms. In the
top panel we plot the effects of the pulse broadening on a simulated
Gaussian pulse from a pulsar with a spin period of 50 ms and a pulse
width of 10 per cent.  The dotted line shows the original pulse
profile with no acceleration effects. The thick black line shows the
full effects of the pulse broadening caused by acceleration and the
thin black line shows the `acceleration-corrected' profile.
\begin{figure}
\includegraphics[scale=0.62]{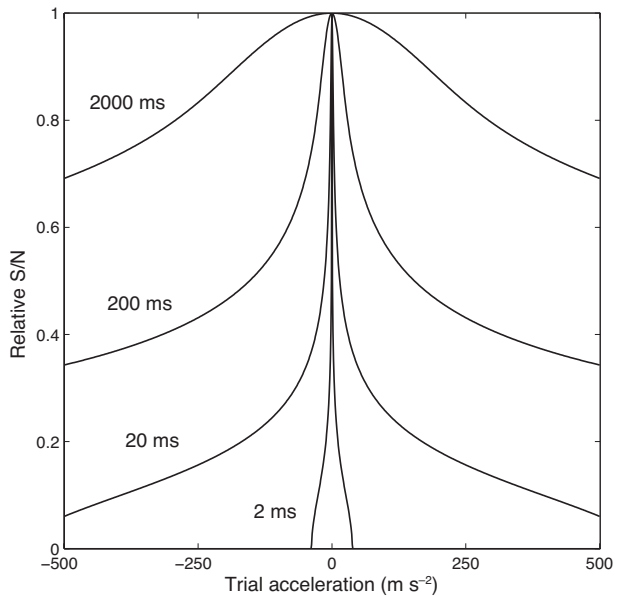}
\caption{Relative pulse S/N variation with incorrect
acceleration trial for a range of pulsar periods. In all cases the
pulse duty cycle is five per cent, and the signal has a true
acceleration of $0\;{\rm m\;s^{-2}}$.}
\label{f:accel_range}
\end{figure}
The reduction in pulse amplitude, caused by broadening, depends on the
intrinsic width of the pulse i.e. assuming an constant duty cycle,
shorter period pulsars will experience increased broadening compared
to longer period pulsars. In Figure~\ref{f:accel_range} we show how
the pulse S/N (related to the effective width and period of the pulse,
see e.g. \citeauthor{lk05}~(2005) page 129) changes with incorrect
acceleration trial for a number of pulse periods, ranging from two
milliseconds to two seconds.  In these examples the pulse duty cycle
is fixed at five per cent and the actual acceleration of the signal is
$0\;{\rm m\;s^{-2}}$. Shorter period pulsars are smeared by incorrect
acceleration trials more quickly than their longer period
counterparts. As well as showing the importance of correcting for
pulse broadening caused by constant accelerations, the curves
displayed in Figure~\ref{f:accel_range} provide a useful diagnostic
for genuinely accelerated pulsar signals.

\begin{figure}
\includegraphics[scale=0.51]{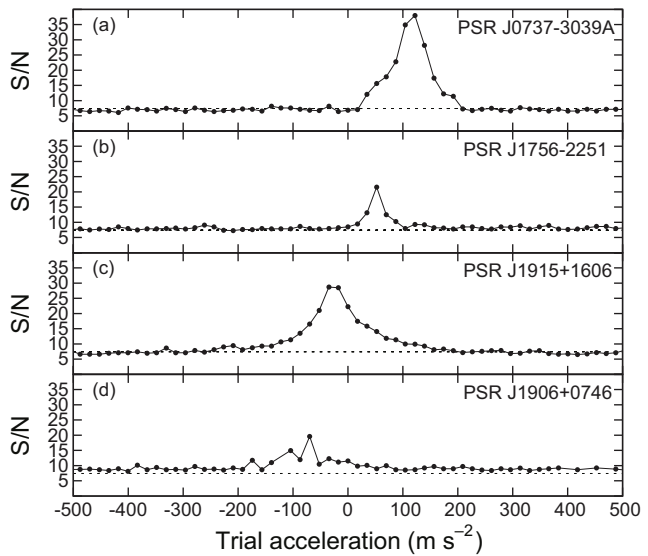}
\caption{Spectral S/N as a function of trial acceleration for four
  known relativistic binary pulsars using the {\sc PMACCN} search
  algorithm. The horizontal dotted lines (at ${\rm S/N}=7.4$)
    show the minimum value of S/N considered not to be from the
    effects of statistical noise, based on the false-alarm
    probability (e.g. \citeauthor{lk05}~2005).}
\label{f:accncurves}
\end{figure}

\subsection{Algorithm tests}
\label{s:algorithm_tests}
The {\sc PMACCN} acceleration search algorithm has been tested on
observations of a number of known relativistic binary pulsars, and using
data from simulations of compact DNS and NS-BH systems
(\citeauthor{Eatough:thesis}~2009). Post facto tests of the algorithm,
described in Section~\ref{s:add_algorithm_tests}, have also been performed. The
results from tests on real observations of relativistic binary pulsars
are presented in Figure~\ref{f:accncurves}.  Each panel shows the
spectral S/N, from the fluctuation spectrum, as a function of trial
acceleration for 1050~s integrations with the PMPS observing
system. PSR A in the Double Pulsar system is shown in panel (a) and is
found at an acceleration of 122 m s$^{-2}$. Panel (b) shows PSR
J1756$-$2251, detected at an acceleration of 52 m s$^{-2}$. In panel
(c) the binary pulsar, B1913$+$16 is found with an acceleration of
$-$26 m s$^{-2}$. Finally panel (d) shows the binary pulsar
J1906$+$0746 detected at $-$70 m s$^{-2}$.  Standard non-accelerated
pulsar searches would give spectral S/Ns roughly equal to the value in
the 0 m s$^{-2}$ acceleration trial. In these tests the PMACCN
acceleration search increases the spectral S/Ns by a factor of between
at least two and five compared to a standard non-accelerated pulsar
search. However, this improvement is dependent upon the orbital phase
at which the pulsar was observed and is discussed in more detail in
Section~6.

\begin{figure*}
\includegraphics[scale=0.45]{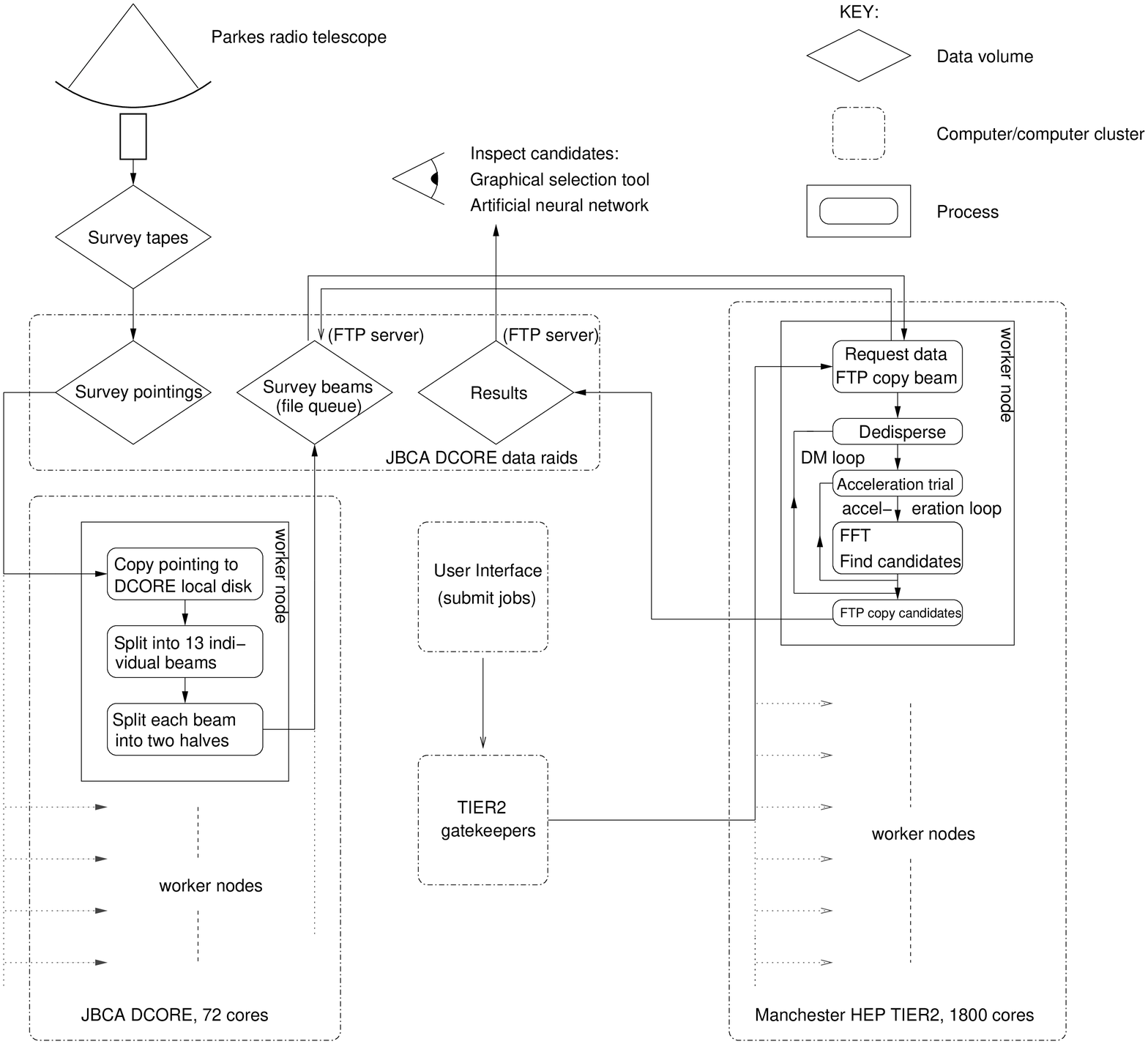}
\caption{Flow diagram summarizing the PMACCN data processing using the
  TIER2 facility. The survey pointings were stored on an archive RAID
  at the Jodrell Bank Centre for Astrophysics (JBCA). Each pointing
  was separated into 26 half integrations using the DCORE cluster and
  stored for processing. Jobs were submitted to the TIER2 facility via
  the `user interface', a desktop machine at the JBCA. Once a
  processing job started on a worker node a request for data was made
  to a control program running on the JBCA RAID host
  computer. Available files were then copied and processed on the
  worker nodes. The candidates from the search were sent back to the
  DCORE cluster where they are viewed with either a graphical
  selection tool or an artificial neural network
  (\citeauthor{emk+10}~2010).}
\label{f:dataflow}
\end{figure*}

\section{The Processing}
\label{s:grid_processing}
It was estimated that processing the PMPS using the acceleration
algorithm described in Section~3 would require an order of
magnitude more computational operations than the
\citeauthor{fsk+04}~(2004) processing used. An opportunity arose to
use the computing facilities of The University of Manchester, High
Energy Physics (HEP) group. Their TIER2 computing
facility\footnote{\footnotesize
  http://www.hep.man.ac.uk/computing/tier2} consists of 900 dual-core
nodes that form part of a network of high-performance computers for
the analysis of particle physics data from the Large Hadron Collider,
known as GridPP\footnote{\footnotesize
  http://www.gridpp.ac.uk}. GridPP was built by a collaboration of 19
UK universities, the Rutherford Appleton Laboratory and Conseil
Europ\'{e}en pour la Recherche Nucl\'{e}aire (CERN). The UK's GridPP
is connected to a much wider global collaboration of computing centres
called the Worldwide LHC Computing Grid\footnote{\footnotesize
  http://lcg.web.cern.ch/LCG/} and in total provides the equivalent
computing power of approximately $200,000$ CPUs.  One of the primary
aims of computational grids, such as GridPP, is to exploit and share
the computational power available from disparate sets of computers
or computer clusters that are connected by the Internet.  Such
`grid' based distributed computing lends itself well to the
highly-parallel nature of processing the data from pulsar surveys.

\begin{figure*}
\begin{center}
\hrule
\vspace{0.2truecm}
\includegraphics[scale=0.66]{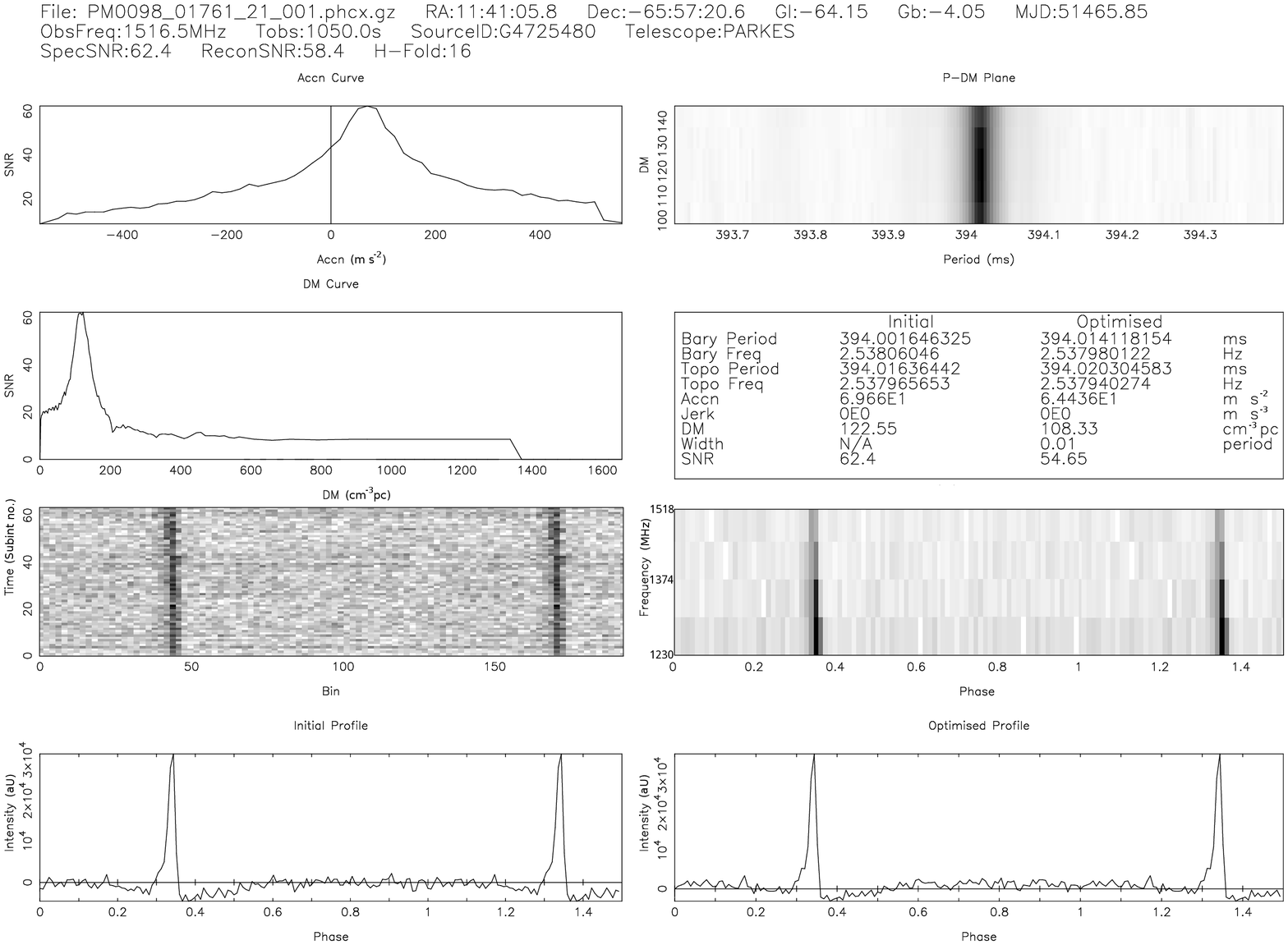}
\vspace{0.2truecm}
\hrule
\caption{Example candidate plot from the PMACCN
  algorithm generated for visual inspection of the quality of pulsar
  candidates and showing a detection the relativistic binary pulsar,
  J1141$-$6545. Starting clockwise from the bottom left the candidate
  plot displays: the integrated pulse profile, folded at the initial
  spectral discovery period and dispersion measure (DM); 64 temporal
  subintegrations of the observation, showing how the pulse varies
  with time; the spectral S/N as a function of a wide range of trial
  DMs (the DM-S/N curve); the spectral S/N as a function of trial
  acceleration (the acceleration-S/N curve: here the pulsar has been
  detected with an acceleration of 64.4 ${\rm ms}^{-2}$); the
  period-DM diagram, that shows how the S/N varies with small changes
  in the folding period and DM; the basic discovery parameters before
  and after time domain optimisation; stacked pulses across four
  frequency subbands, showing how the pulse varies with observing
  frequency, and finally the integrated pulse profile, folded at the
  optimum period and dispersion measure. The dips, visible at the
  edges of the integrated pulse, are caused by application of the
  zero-DM filter (\citeauthor{ekl09}~2009).}
\label{f:1141}
\end{center}
\end{figure*}

\subsection{Data processing overview}
\label{s:proc_overview}
Data processing was divided into two main stages. Firstly `local'
processing, where the PMPS pointings were separated into 13 individual
beams and each integration divided into two halves on the Jodrell Bank
Centre for Astrophysics (JBCA) DCORE computer cluster.  This was
followed by `external' beam processing, where each half integration
was acceleration-searched on the TIER2 facility. The flow of data and
processing steps from acquisition at the telescope up to the
inspection of pulsar candidates generated from the search is depicted
in Figure~\ref{f:dataflow}.

In summary, the PMPS data were stored in pointings of 13 independent
beams on a RAID system at the JBCA.  Each data file consists of
$2^{23}$ single-bit samples for each of the 96 3-MHz-wide frequency
channels. The raw data were first separated into 13 individual beams
using a dedicated \textsc{fortran} program, \textsc{sc\_td}. Each beam
was then converted into multi-channel \textsc{sigproc} format, for use
with the \textsc{sigproc} pulsar processing
suite\footnote{\footnotesize http://sigproc.sourceforge.net/}, and
split into two 1050~s segments. Each of the half-length integrations
was stored on the JBCA RAID system, ready for processing on the TIER2
facility. These initial processing tasks were performed on the DCORE
computer cluster, to avoid excessive levels of data transfer to the
TIER2 facility.

The submission of processing jobs to the TIER2 facility was controlled
by the `user interface' to GridPP: a linux desktop machine located at
the JBCA. Processing jobs were submitted to the TIER2 `gatekeepers';
machines that maintained a batch queue of jobs.  Once a processing job
started on a TIER2 worker node, a request was sent to a \textsc{java}
control program, running at the JBCA, that kept a queue of
un-processed data files. When a data file became available, it was
copied to the TIER2 worker node. The data were then searched for
periodic signals following the standard search procedures outlined in
\citeauthor{mlc+01}~(2001) and \citeauthor{lk05}~(2005): Firstly, the
96 channels were dedispersed to four frequency sub-bands at each value
of trial dispersion measure (DM). Just prior to dedispersion the
harmful effects of radio frequency interference (RFI) were reduced by
application of the zero-DM filter (\citeauthor{ekl09}~2009). These
sub-band data were stored for folding and optimisation of the
candidates generated from the acceleration search. The subbands were
then dedispersed (without re-application of the zero-DM filter) to a
single channel timeseries at the same trial DM. Every four sequential
samples in this time series were then summed together, to give a 1 ms
sample interval before performing the coherent acceleration search as
outlined in Section 3. Features above a pre-defined threshold S/N
(calculated from the expected false-alarm probability) in the
resulting fluctuation spectra were stored in a candidate list along
with the associated DM and acceleration. At this point `reconstructed
S/Ns' were also computed and stored\footnote{Reconstructed S/N is
  the S/N of a pulse profile generated from the inverse Fourier
  transform of the complex frequency components of a signal
  (e.g. Appendix C, \citeauthor{fau04}~2004,
  \citeauthor{lk05}~2005).}. Harmonic summation of two, four, eight
and 16 harmonics was performed to provide sensitivity to narrow pulses
(\citeauthor{th69}~1969).  The threshold S/N was reduced in successive
harmonic folds as the probability density function of spectral
features evolves from a Rayleigh to a Gaussian distribution. The whole
process was repeated for each trial value of DM and acceleration.

After dispersion and acceleration searching was complete the candidate
list was checked for the most significant signals (sorted by their
reconstructed S/N) at each DM and acceleration. The DM and
acceleration for which each signal was strongest was noted and any
harmonics of the signal were checked for and ignored if identified.
Signals down to a threshold ${\rm S/N}\sim9$, from each of the
1050 s segments, were stored in a final candidate list. Candidates in
this list were then `optimised' in the time domain using the {\sc
  pulsarhunter} software package\footnote{\footnotesize
  http://sourceforge.net/projects/pulsarhunter/}. The appropriate
sub-band data file was divided into 64 temporal sub-integrations after
which searches over narrow ranges in period, DM and acceleration were
performed to find the best integrated pulse S/N. The results of this
search and the initial acceleration search were placed into `candidate
plots' for viewing. An example of one such candidate plot, covering
the position of PSR J1141$-$6545, is visible in
Figure~\ref{f:1141}. Finally, all results were compressed with {\sc
  gzip} and returned to the JBCA for inspection.

\subsection{Post-processing overview}
\label{s:post_proc_overview}
The total number of pulsar candidates produced by reprocessing the
PMPS with the PMACCN search algorithm was in the region of 16 million;
a number far in excess of previous analyses. As first noted by
\citeauthor{fsk+04}~(2004), such large numbers of pulsar candidates
can, in practice, only be inspected with the aid of a graphical
selection tool. For this purpose the graphical selection tool
\textsc{jreaper} (\citeauthor{kel+09}~2009) was employed.  Despite the
advanced filtering techniques that can be performed with
\textsc{jreaper} the number of candidates requiring visual inspection
still remained large. As such, an automatic candidate inspection
technique, utilising an artificial neural network (ANN), was developed
to complement the manual searches performed with \textsc{jreaper}
(\citeauthor{emk+10}~2010).

Promising pulsar candidates selected with either \textsc{jreaper} or
by the ANN were placed into three categories for follow-up
observation: class one, for candidates that had a large probability of
being new pulsar discoveries, class three for candidates that had a
smaller chance of being new pulsars, and class two, for candidates
that were in between the former categories.  The majority of
candidates viewed were non-pulsar signals caused by either RFI or
noise fluctuations. These signals were discounted.

\section{Discovery and timing of 16 pulsars}
\label{s:discoveries}
Reprocessing of the PMPS with the \textsc{PMACCN} algorithm has
yielded 16 new pulsar discoveries. Despite the improved acceleration
search algorithm used in this analysis, no previously unknown
relativistic binary pulsars have been found. However, all known
relativistic binary pulsars in the survey region have been re-detected
with a higher significance than obtained in previous analyses. A
number of promising binary candidates have been identified, but to
date none have been re-detected despite follow-up observations (see
Section~6.2).

Not all of the pulsars found in this work have been detected here for
the first time. During comparison of the candidates generated from
this search with those from the previous \citeauthor{fsk+04}~(2004)
analysis, it was noticed that a number of the previous class one
candidates had never been re-observed. These oversights were
undoubtedly due to the large number of genuine pulsar discoveries
making tracking of the status of each candidate a difficult task. As
part of this work, confirmation observations of these sources were
performed after which they were added to timing programmes at either
the Parkes or Jodrell Bank Observatories. All the pulsars have been
timed for a period long enough to determine an accurate position and
to derive the basic pulsar parameters. All timing measurements were
performed with the PSRTIME
software\footnote{http://www.jb.man.ac.uk/$\sim$pulsar/observing/progs/psrtime.html}.

Table~2 presents the pulsar name, the J2000 right ascension and
declination from the timing solution, the Galactic coordinates, the
S/N of the discovery observation after the time-domain folding and
optimisation, the candidate selection tool by which the pulsar was
discovered, the mean flux density at $\sim$1.4~GHz from the survey
detection, $S_{1400}$, and the pulse width at 50 per cent of the peak
of the mean pulse profile, $W_{50}$. Uncertainties on the values
presented, where relevant, are given in parentheses and refer to the
last digit. Flux densities have been corrected for positional offsets
in the survey detection. Sources marked with a {\bf *} were not
detected in this analysis but were found by reobservation of the
unconfirmed class one candidates from the previous
\citeauthor{fsk+04}~(2004) reprocessing. Sources marked with a
$^{\clubsuit}$ were detected both in this analysis and the previous
analysis but only confirmed now. Sources with no markings have been
detected here for the first time. In summary, five of the pulsars were
not detected in this reprocessing, three were detected in both, and
seven have been newly discovered. The discoveries of PSRs
J1539$-$4828, J1819$-$1717, and J1835$-$0114 have already been
mentioned in \citeauthor{ekl09}~(2009) and the discovery of PSR
J1926$+$0737, using an ANN, is described in
\citeauthor{emk+10}~(2010).

\begin{figure*}
\begin{center}
\includegraphics[scale=0.6]{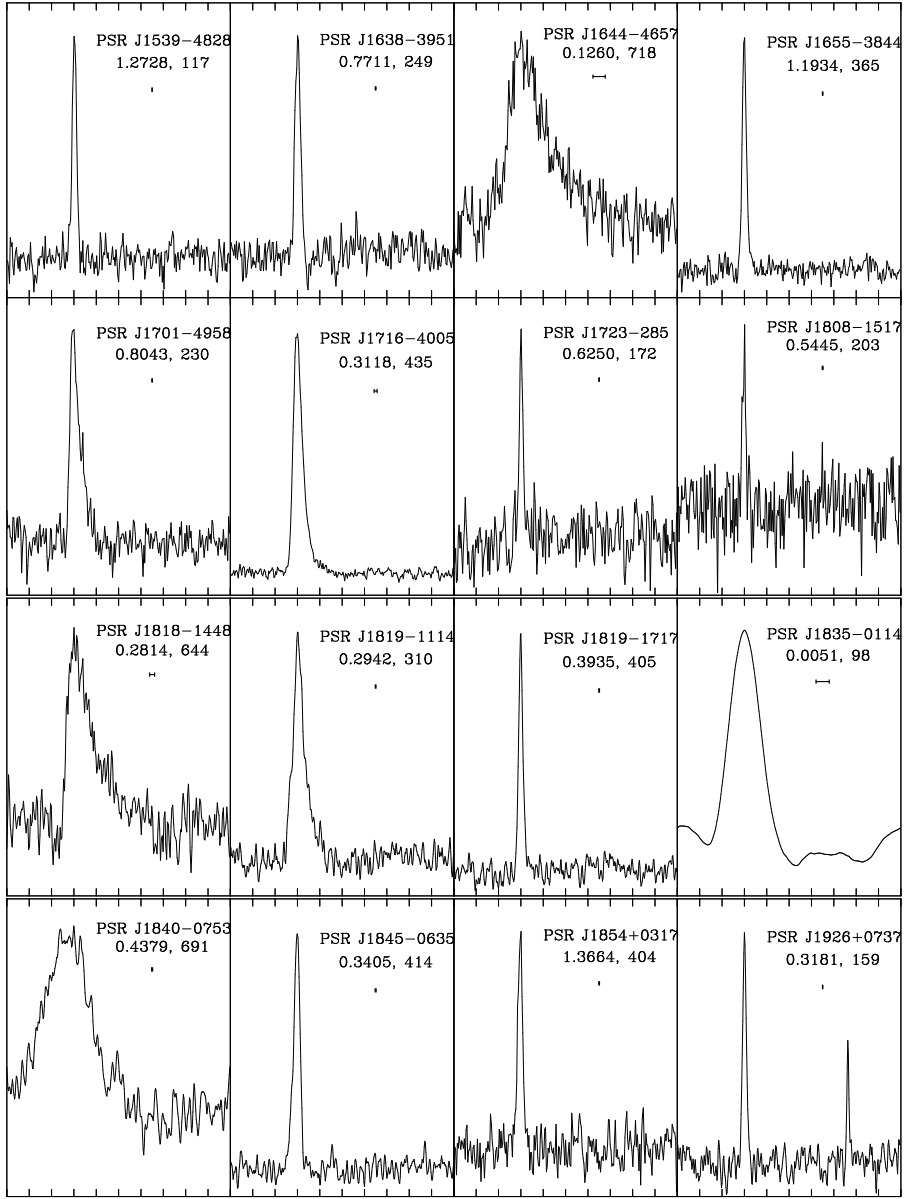}
\caption {Mean pulse profiles over a phase of zero to one at 1.4
  GHz for the 16 pulsars. The peak of each profile has been centered
  at a phase of 0.3. Each panel also displays the pulsar name, pulse
  period in seconds, and dispersion measure in
    cm$^{-3}$pc. The effective time resolution of the profile,
  including the effects of interstellar dispersion, is indicated by
  the horizontal error bar below the text.}
\label{f:profiles}
\end{center}
\end{figure*}

Table~3 gives barycentric pulse periods, period derivatives, epoch of
the period, the number of pulse time of arrival (TOA) measurements in
the timing solution, the MJD range of the timing observations, the
final rms residuals of the timing model, and the dispersion
measure. Table~4 gives the following derived parameters of the 16
pulsars: characteristic age, $\tau_{c}$, surface magnetic field,
$B_{s}$, rate of loss of rotational energy, $\dot{E}$, the dispersion
derived distance using the NE2001 electron density distribution model
(\citeauthor{cl02}~2002), the corresponding height above the Galactic
plane, $z$, and finally the inferred luminosity at 1400 MHz,
$L_{1400}$. High S/N mean pulse profiles of each of the 16 pulsars at
1374 MHz are presented in Figure~\ref{f:profiles}.

Excluding the pulsars that did not appear in this processing
(i.e. sources marked with {\bf *}), the remaining sample exhibits a
wide range in flux densities.  The discovery of PSR J1716$-$4005, a
source with a large flux density that was completely undetected in
previous analyses, has most likely been enabled by effective removal
of RFI, by use of the zero-DM filter, and improved candidate selection
tools. A number of the sources are below the limiting flux density of
our search ($S_{1400}\sim0.2$ mJy for the centre of the central beam)
which is estimated to be of order $\sqrt 2$ times less sensitive than
that in the original searches of the PMPS because of the use of half
length, 1050 s integrations.  The zero-DM filter may account for this
as its use can improve the limiting survey sensitivity by the
effective removal of RFI (\citeauthor{ekl09}~2009). In general the
pulsars are consistent with the sample of pulsars found by the PMPS as
a whole viz. low flux density but high luminosity because the pulsars
are found at relatively large distances.  PSR J1818$-$1448 has a the
largest period derivative of the pulsars discovered implying an age of
just 720 kyr and making it the youngest of the new
sources. Unfortunately, this age is too great to expect any supernova
remnant association to be visible.

\begin{table*}
\label{t:posns}
\caption{\normalsize Positions, survey beam in which the pulsar was
  discovered, candidate selection tool used in the discovery, folded
  S/N, flux densities and pulse widths of the 16 pulsars discovered in
  this work. Positional uncertainties are derived
    from the one-sigma error of the timing model fit obtained using the
    PSRTIME software. See Section 5. for an explanation of the
  symbols.}
\begin{small}
\begin{center}
\begin{tabular}{lllrrrllrr}
\\\hline\hline
\multicolumn{1}{l}{PSR J} & \multicolumn{1}{l}{R.A. (J2000)} & \multicolumn{1}{l}{Dec. (J2000)} & \multicolumn{1}{c}{{\it l}} &
 \multicolumn{1}{c}{\it b} & \multicolumn{1}{c}{S/N} & \multicolumn{1}{c}{Selection} & \multicolumn{1}{c}{$S_{140
0}$} & \multicolumn{1}{c}{$W_{50}$}\\ 
& \multicolumn{1}{c}{(h min s)} & \multicolumn{1}{c}{$(^{\circ}\;^{\prime}\;^{\prime\prime})$} & \multicolumn{1}{c}{($^\circ$)} &
 \multicolumn{1}{c}{($^\circ$)} & \multicolumn{1}{c}{} & \multicolumn{1}{c}{tool} & \multicolumn{1}{c}{(mJy)} & \multicolumn{1}{c}{(ms)} \\\hline
J1539$-$4828               & 15:39:40.84(6)   & $-$48:28:57(1)    & 329.43   & $+$5.54 &  9.5 & {\sc jreaper} & 0.20(2) &  50.9 \\
J1638$-$3951{\bf *}        & 16:38:15.56(5)   & $-$39:51:59(1)    & 343.00   & $+$4.74 & 10.2 & n/a           & 0.21(2) &  61.7 \\
J1644$-$4657{\bf *}        & 16:44:38.5(1)    & $-$46:57:38(4)    & 338.43   & $-$0.81 &  8.9 & n/a           & 0.59(5) &  40.3 \\
J1655$-$3844{\bf *}        & 16:55:38.66(5)   & $-$38:44:09(1)    & 346.04   & $+$2.91 & 10.9 & n/a           & 0.25(2) &  95.5 \\\\
J1701$-$4958$^{\clubsuit}$    & 17:01:12.83(6)   & $-$49:58:33(2)    & 337.84   & $-$4.84 & 11.2 & {\sc jreaper} & 0.26(1) & 128.7 \\
J1716$-$4005               & 17:16:42.06(4)   & $-$40:05:27(2)    & 347.42   & $-$1.15 & 56.2 & {\sc jreaper} & 1.37(1) &  48.9 \\
J1723$-$2852               & 17:23:58.2(3)    & $-$28:52:51(15)   & 357.47   & $+$4.01 & 10.6 & {\sc jreaper} & 0.13(1) &  50.0 \\
J1808$-$1517               & 18:08:39.09(2)   & $-$15:17:40(2)    &  14.47   & $+$2.24 & 12.5 & {\sc jreaper} & 0.35(3) &  21.8 \\\\
J1818$-$1448{\bf *}        & 18:18:27.91(6)   & $-$14:48:38(7)    &  16.03   & $+$0.39 &  9.7 & n/a           & 0.19(1) &  45.0 \\
J1819$-$1114$^{\clubsuit}$    & 18:19:28.78(1)   & $-$11:14:43(1)    &  19.29   & $+$1.86 & 13.7 & {\sc jreaper} & 1.07(5) &  94.1 \\
J1819$-$1717               & 18:19:43.40(5)   & $-$17:17:16(5)    &  13.99   & $-$1.05 & 11.3 & {\sc jreaper} & 0.26(2) &  31.5 \\
J1835$-$0114               & 18:35:21.9179(9) & $-$01:14:33.61(2) &  29.99   & $+$3.01 & 11.8 & {\sc jreaper} & 0.10(1) &   0.2 \\\\
J1840$-$0753               & 18:40:47.6(1)    & $-$07:53:32(6)    &  24.70   & $-$1.24 & 13.1 & {\sc jreaper} & 0.37(2) & 140.1 \\
J1845$-$0635$^{\clubsuit}$    & 18:45:07.406(9)  & $-$06:35:23.4(8)  &  26.35   & $-$1.60 & 31.0 & {\sc jreaper} & 0.36(1) &  27.2 \\
J1854$+$0317{\bf *}        & 18:54:29.06(7)   & $+$03:17:31(3)    &  36.21   & $+$0.82 & 10.5 & n/a           & 0.12(1) & 109.3 \\
J1926$+$0737               & 19:26:33.740(6)  & $+$07:37:07.2(3)  &  43.74   & $-$4.26 & 10.2 & {\sc ann}     & 0.11(1) &  25.4 \\
\hline
\end{tabular}
\end{center}
\end{small}
\end{table*}

\begin{table*}
\label{t:timingdata}
\caption{\normalsize Timing parameters and dispersion measures of the
  16 pulsars discovered in this work. Spin period
    uncertainties are derived from the one-sigma error of the timing model
    fit obtained using the PSRTIME software. See Section~5. for an explanation of the
  symbols.}
\begin{small}
\begin{center}
\begin{tabular}{lllcrcrr}
\\\hline\hline
\multicolumn{1}{l}{PSR J} & \multicolumn{1}{c}{Period, {\it P}} & \multicolumn{1}{c}{$\dot{P}$} & \multicolumn{1}{c}{Epoch} & \multicolumn{1}{c}{$N_{\rm toa}$} & \multicolumn{1}{c}{Data span} & \multicolumn{1}{c}{Residual} & \multicolumn{1}{c}{DM}\\
& \multicolumn{1}{c}{(s)} & \multicolumn{1}{c}{$(10^{-15})$} & \multicolumn{1}{c}{(MJD)} & & \multicolumn{1}{c}{MJD} & \multicolumn{1}{c}{$\mu {\rm s}$} & \multicolumn{1}{c}{$\rm cm^{-3} pc$}\\\hline
J1539$-$4828               & 1.27284159978(14)     & 1.267(2)     & 54845.0 &  22 & 54647$-$55043 &  1590 & 117(5)\\
J1638$-$3951{\bf *}        & 0.77113024025(8)      & 0.59(4)     & 54863.3 &  22 & 54683$-$55043 &  1014 & 249(4)\\
J1644$-$4657{\bf *}        & 0.12596223249(4)      & 0.75(2)     & 54835.1 &  16 & 54683$-$54988 &  1792 & 718(6)\\
J1655$-$3844{\bf *}        & 1.19343919921(14)     & 2.00(6)     & 54863.3 &  22 & 54683$-$55043 &  1144 & 365(11)\\\\
J1701$-$4958$^{\clubsuit}$    & 0.80430423739(12)     & 0.11(4)     & 54844.6 &  25 & 54646$-$55043 &  2332 & 230(5)\\
J1716$-$4005               & 0.31181276456(4)      & 2.87(4)     & 54942.8 &  15 & 54842$-$55043 &   160 & 435(3)\\
J1723$-$2852               & 0.6250339127(12)      & 0.9(16)     & 54948.6 &  31 & 54862$-$55691 &  1895 & 172(6)\\
J1808$-$1517               & 0.544549207067(8)     & 2.6671(5)   & 55105.4 &  94 & 54596$-$55615 &  1035 & 205(5)\\\\
J1818$-$1448{\bf *}        & 0.28137130037(6)      & 6.15(2)     & 54863.1 &  28 & 54683$-$55043 &  3142 & 644(6)\\
J1819$-$1114$^{\clubsuit}$    & 0.29416251457(2)      & 0.566(6)    & 54849.9 &  81 & 54652$-$55048 &  1222 & 310(5)\\
J1819$-$1717               & 0.39352163764(6)      & 3.455(6)    & 54756.2 &  75 & 54529$-$54983 &  1698 & 405(2)\\
J1835$-$0114               & 0.005116387644239(12) & 0.000007(4) & 54786.8 & 142 & 54540$-$56010 &   107 &  98(1)\\\\
J1840$-$0753               & 0.4378690017(2)       & 0.10(6)     & 54803.9 & 140 & 54559$-$55049 & 20061 & 691(7)\\
J1845$-$0635$^{\clubsuit}$    & 0.34052764699(2)      & 4.491(8)    & 54844.5 &  65 & 54656$-$55033 &   744 & 414(1)\\
J1854$+$0317{\bf *}        & 1.3664496471(2)       & 1.85(16)    & 54863.5 &  18 & 54683$-$55043 &  2472 & 404(8)\\
J1926$+$0737               & 0.318062050971(12)    & 0.375(4)    & 54833.6 &  48 & 54619$-$55048 &   704 & 159(2)\\
\hline
\end{tabular}
\end{center}
\end{small}
\end{table*}

\begin{table*}
\label{t:derived_params}
\caption{\normalsize Derived parameters of the 16 pulsars discovered
  in this work. Distances have been calculated based on the
  \citeauthor{cl02}~(2002) free electron density model of the Galaxy.
  See Section 5 for an explanation of the symbols.}
\begin{small}
\begin{center}
\begin{tabular}{lrrrrrr}
\\\hline\hline
\multicolumn{1}{l}{PSR J} & \multicolumn{1}{c}{$\tau_{c}$} & log$B_{s}$ & \multicolumn{1}{c}{log$\dot{E}$} & \multicolumn{1}{c}
{Distance} &  \multicolumn{1}{c}{z} & \multicolumn{1}{c}{$L_{1400}$}\\
& \multicolumn{1}{c}{(Myrs)} & \multicolumn{1}{c}{(G)} & \multicolumn{1}{c}{(erg $\rm s^{-1}$)} & \multicolumn{1}{c}{(kpc)} & \multicolumn{1}{c}{(kpc)} & \multicolumn{1}{c}{(mJy kpc$^2$)}\\\hline 
J1539$-$4828               &    15.96 & 12.10 & 31.38 &  3.6 & $+$0.35 &  1.3\\
J1638$-$3951{\bf *}        &    20.65 & 11.83 & 31.71 &  5.4 & $+$0.45 &  3.2\\
J1644$-$4657{\bf *}        &     2.65 & 11.49 & 34.17 &  8.2 & $-$0.12 & 14.8\\
J1655$-$3844{\bf *}        &     9.43 & 12.19 & 31.67 &  6.5 & $+$0.33 &  5.1\\\\
J1701$-$4958$^{\clubsuit}$    &   115.53 & 11.47 & 30.92 &  4.9 & $-$0.41 &  5.8\\
J1716$-$4005               &     1.72 & 11.98 & 33.57 &  5.4 & $-$0.11 & 33.0\\
J1723$-$2852               &    10.97 & 11.88 & 32.16 &  3.4 & $+$0.24 &  1.4\\
J1808$-$1517               &     3.23 & 12.08 & 32.81 &  3.8 & $+$0.15 &  1.6\\\\
J1818$-$1448{\bf *}        &     0.72 & 12.12 & 34.04 &  7.4 & $+$0.05 &  8.2\\
J1819$-$1114$^{\clubsuit}$    &     8.21 & 11.61 & 32.94 &  5.0 & $+$0.16 &  9.8\\
J1819$-$1717               &     1.80 & 12.07 & 33.35 &  5.6 & $-$0.10 &  5.3\\
J1835$-$0114               & 11548.42 &  8.28 & 33.31 &  2.7 & $+$0.14 &  0.7\\\\
J1840$-$0753               &    69.18 & 11.32 & 31.67 & 10.0 & $-$0.22 & 30.0\\
J1845$-$0635$^{\clubsuit}$    &     1.20 & 12.09 & 33.65 &  7.1 & $-$0.20 & 17.6\\
J1854$+$0317{\bf *}        &    11.67 & 12.20 & 31.46 &  7.5 & $+$0.11 &  6.2\\
J1926$+$0737               &    13.40 & 11.54 & 32.66 &  5.6 & $-$0.42 &  3.1\\
\hline
\end{tabular}
\end{center}
\end{small}
\end{table*}

\subsection{Notable discoveries}
\label{s:notable_discs}

\subsubsection{PSR J1835$-$0114, a binary millisecond pulsar}
\label{s:1835}
Soon after the discovery of PSR J1835$-$0114, it was noticed that the
spin period was not constant but changing periodically. As no sign of
period variation was measured in the search output, i.e. over 1050 s
integration, the behaviour was thought to be typical of that of a
pulsar in a long orbital period binary system. Continued timing
observations at Jodrell Bank Observatory have shown that the pulsar is
in a 6.7-day orbit, probably around a low mass companion. The measured
binary parameters of this system are given in Table~5. The pulsar is
in a circular orbit ($e\sim1\times10^{-5}$), typical of pulsars in
binary systems that have undergone a long period of accretion powered
spin up (Phinney 1992)\nocite{phi92}. In non-relativistic systems
estimation of the mass of the pulsar and companion, $m_{p}$ and
$m_{c}$, can only be made through accurate measurements of the mass
function given by
\begin{equation}
\label{e:massfunc}
f(m_{p},m_{c},i) = \frac{(m_{c}\,{\rm sin}\,i)^3}{(m_{p}+m_{c})^2} = \frac{4\pi^2}{G}\frac{a^3{\rm sin}^3 i}{P
_{b}^2} = 0.00241709 M_{\odot} ,
\end{equation}
where $a$ is the semi-major axis of the orbit, $i$ is the inclination
angle and the other terms have their usual meanings. In order to
estimate the mass of the companion a typical neutron star mass of 1.35
$M_{\odot}$, from the known neutron star masses measured in
relativistic binaries (Thorsett \& Chakrabaty 1999)\nocite{tc99}, has
been assumed.  Random orbital axis orientation arguments allow us to
assume a median inclination angle of 60$^{\circ}$, see
e.g. \cite{ls06}, page 127. With the observed value of the mass
function, an iterative solution for $m_{c}$ in
Equation~\ref{e:massfunc} gives a most likely companion mass of 0.21
$M_{\odot}$; indicative of a white dwarf star.

\subsubsection{PSR J1808-1517, an intermittent pulsar}
\label{s:1808}
\begin{figure}
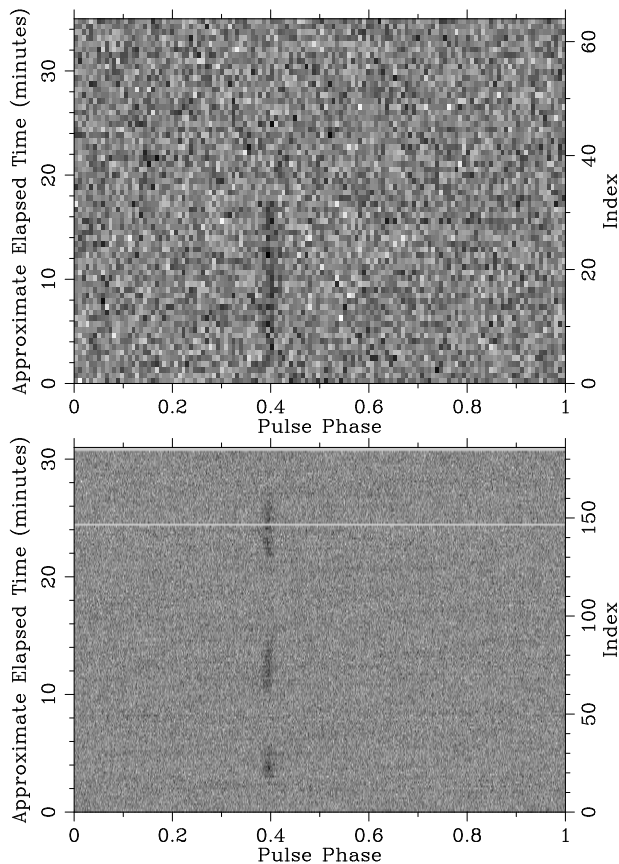

\includegraphics[scale=0.35,angle=-90]{figs/pmps_obs.ps}
\includegraphics[scale=0.35,angle=-90]{figs/eff_obs.ps}
\caption{Phase resolved pulse amplitude as a function of time for
the original PMPS observation of PSR J1808-1517 (top panel) and a
re-observation performed at the Effelsberg observatory (bottom
panel). Subintegrations are plotted every $\sim 30$ s and $\sim12$ s
for the PMPS and Effelsberg observations respectively.}
\label{f:1809_eff}
\end{figure}

PSR J1808$-$1517 was a class one candidate that was detected in only
one half of the original 35 minute observation. Analysis of the full
integration, revealed the pulsed signal disappearing approximately
half way through the observation (see Figure~\ref{f:1809_eff} top
panel). The candidate class, combined with observed parameters typical
of normal pulsars ($P\sim544$ ms, DM $\sim205$~cm$^{-3}$pc, pulse duty
cycle $\sim$4 per cent) pointed to a clear pulsar detection. However,
in approximately seven hours of follow up observations performed on
different dates at the Parkes observatory and regular observations
from March 13th to September 9th, 2008 at the Jodrell Bank
observatory, no re-detection was made. Despite the expected difficulty
in confirmation it was decided to continue with observations at the
Parkes observatory at approximately monthly intervals. In March~2010 a
re-detection was made. To confirm the intermittent nature of this
pulsar, re-observations were performed at the Effelsberg observatory.
The bottom panel of Figure~\ref{f:1809_eff} shows temporal
subintegrations over a $31$ minute observation done at
Effelsberg. During this observation the pulsar exhibits nulls lasting
the order of a few minutes, with a nulling fraction of $\sim 50$
per cent. Because of the earlier non-detections, we suggest the
  length of nulls vary and could be at least as long as a typical
  observation time ($\gtrsim 30$ minutes). Based on the Effelsberg
observation the switching time scale between states is $\lesssim 12$ s
($\lesssim 23$ pulse periods).

This analysis of the PMPS, that has searched independent halves of the
original 35 minute integrations, has aided in the detection of
this pulsar. The addition of the second half of the original
observation, when the pulsar was not visible, reduces the integrated
pulse S/N down to 8.5 and places the pulsar on the limit of previous
detection thresholds.

\subsubsection{J1926$+$0737, a pulsar with an interpulse}
\label{s:1926}
PSR J1926$+$0737, that was found using an ANN
(\citeauthor{emk+10}~2010), exhibits an interpulse at a rotational
phase $\sim \,$180$^{\circ}$ from the main pulse.  This characteristic
implies that the angle between the magnetic axis and the rotational
axis, $\alpha$, is $\sim \,$90$^{\circ}$ and that the pulsar must be
an orthogonal rotator, where emission from both magnetic poles is
visible. This pulsar adds to the population of 31 normal pulsars (with
$P>20$ ms) that exhibit interpulses
(\citeauthor{2011MNRAS.414.1314M}~2011). The properties of pulsars
with interpulses are of importance for understanding the evolution of
pulsar beams, which has implications in the long term spin-down
behaviour of pulsars and also in the number of pulsars likely to be
detected in future pulsar surveys
(e.g. \citeauthor{2008MNRAS.387.1755W}~2008,
\citeauthor{2011MNRAS.414.1314M}~2011).

\begin{table}
\label{t:1835_params}
\caption{Binary parameters of PSR J1835$-$0114. Uncertainties on the
  values presented are given in parentheses and refer to the last
  digit.}  
\centering
\begin{small}
\begin{tabular}{lc}\\\hline\hline
Parameter & Value\\\hline
Orbital period (days) & 6.6925427(4)\\
Projected semimajor axis $x=(a/c)$sin$i$ (s) & 4.654423(7)\\
Orbital eccentricity $e$ ($\times 10^{-5}$) & 1.1(3)\\
Longitude of periastron $\omega$ ($^\circ$) & -3(1)\\
Epoch of periastron (MJD) & 54540.7(3)\\
Mass function ($M_{\odot}$) & 0.00241709(4)\\\hline
\end{tabular}
\end{small}
\end{table}

\section{Discussion}
\label{s:discussion}
In this section we start with a brief comparison of our results with
the available predicted number of relativistic binary pulsars in the
PMPS (Section~\ref{s:binary_comparison}). We then outline some of the
possible reasons for the non-detection of new relativistic binary
systems (Section~\ref{s:rbpsr_non_detection}), before describing
further tests of the acceleration algorithm used here
(Section~\ref{s:add_algorithm_tests}). Lastly, Section \ref{s:nonconf}
is a brief discussion of the promising pulsar candidates detected with
PMACCN, but which have never been confirmed in re-observations.

\subsection{Comparison of results to binary population predictions}
\label{s:binary_comparison}
To date, this acceleration search of the PMPS has not increased the
number of relativistic binary systems discovered in the survey.  Prior
to the commencement of this work, and using a method based on the
observed sample of DNS systems
(\citeauthor{2003ApJ...584..985K}~2003),
\citeauthor{2004ApJ...601L.179K}~(2004) made a prediction that
approximately another $4\pm2$ (error estimated from Figure 2. of
\citeauthor{2004ApJ...601L.179K}~(2004)) compact binary pulsars (with
properties similar to either PSRs B1913$+$16, B1534$+$12, and
J0737$-$3039A/B) remained to be discovered in the PMPS data, and would
be detectable with effective acceleration searches. The
\citeauthor{fsk+04}~(2004) re-analysis of the PMPS proved that there
were indeed binary systems that had previously been missed because
acceleration searches were not used, including one compact DNS system,
PSR J1756$-$2251. PSR J1906$+$0746 was also detected with high
significance in this analysis but missed at the candidate selection
stage (\citeauthor{2006ApJ...640..428L}~2006). Currently PSR
J1906$+$0746 is believed to be a young pulsar with a heavy white dwarf
companion, so does not belong to the category of systems described by
Kalogera et al. (2004). The number of relativistic DNS systems
discovered in the PMPS is therefore less than the
\citeauthor{2004ApJ...601L.179K}~(2004) prediction. If PSR
J1906$+$0746 is in fact a DNS system, the current number of compact
DNS in the PMPS would be two; still less than predicted. Whatever the
nature of the companion, the youth of PSR J1906$+$0746 implies that
the underlying population of this type of system could be larger
(\citeauthor{2006ApJ...640..428L}~2006).
  
Alternative Galactic population predictions, from binary synthesis
methods, differ by orders of magnitude in the number of DNS and NS-BH
systems produced because of large uncertainties in the model
parameters used to describe key stages of binary evolution; such as
the magnitude of supernova kicks and the outcome of common envelope
evolution (see e.g. \citeauthor{Bethe:1998}~1998,
\citeauthor{PortegiesZwart:1998}~1998,
\citeauthor{Nelemans:2001}~2001, \citeauthor{Sipior:2002}~2002,
\citeauthor{Belczynski:2002}~2002, \citeauthor{Voss:2003}~2002,
\citeauthor{Sipior:2004}~2004, \citeauthor{Lipunov:2005}~2005,
\citeauthor{Pfahl:2005}~2005, \citeauthor{bbb+10}~2010,
\citeauthor{obg+11}~2011). The population constraints provided by the
binary pulsar search analysis performed here, and similar binary
searches, can be used to improve and calibrate binary population
synthesis codes. However, the result of a non-detection should be
treated with caution due to a number of selection effects and stages
at which pulsar candidates can easily be lost. Some of these points
are discussed in the following section.

\subsection{Possible reasons for the non-detection of new relativistic binaries} 
\label{s:rbpsr_non_detection}

Although this search can help to constrain the numbers of new
relativistic binary pulsar systems, our non-detection does not
necessarily imply that there are no more of these systems to be
found. A more likely scenario is that these pulsars have been missed
because of any one of a number possible reasons, some of which are
listed below. For example, such pulsar systems might be

\begin{itemize}

\item below the limiting sensitivity of the PMPS. For this search of
  1050 s segments, the best sensitivity of $\sim 0.2$~mJy is achieved
  in the centre of the central beam.

\item at an orbital phase where the acceleration was not constant (see
  Section~\ref{s:int_length_orb_phase}).

\item outside the range of orbital accelerations searched (see
  Section~\ref{s:acc_search_range}).

\item undetected because they are in a survey beam with excessive
  RFI. Despite the use of the zero-DM filter, which reduces the amount
  of the fluctuation spectrum that is removed to virtually zero
  (\citeauthor{ekl09}~2009), RFI can reduce the sensitivity of the
  survey. Unfortunately it is difficult to quantify this effect
  (\citeauthor{mlc+01}~2001).

\item detected but not found during candidate selection (see
  e.g. \citeauthor{kel+09}~2009, \citeauthor{emk+10}~2010). Pulsar
  surveys produce many millions of pulsar candidates which are
  typically inspected by eye and which can lead to promising
  candidates being overlooked. The missed discovery of PSR
  J1906$+$0746 in the PMPS illustrates this
  (\citeauthor{2006ApJ...640..428L}~2006). Attempts in this work were
  made to automate this process using an ANN (see Section 4.2 and
  \citeauthor{emk+10}~2010).

\item already detected but cannot be confirmed with follow-up
  observations either because of intermittent emission (see
  e.g. Section \ref{s:1808}) or because the emission beam has
  precessed out of the line-of-sight (see Section~\ref{s:nonconf}).

\item undetected as they were not emitting during the survey
  observation because they are a different class of pulsar,
  e.g. Rotating Radio Transients (\citeauthor{mll+06}~2006) or
  intermittent pulsars (\citeauthor{klo+06}~2006).

\item not in the region of sky covered by PMPS. See below for details
  of the new all-sky pulsar surveys.

\end{itemize}
As already mentioned, some of these points are un-avoidable and others
cannot be addressed until new more sensitive pulsar surveys, such as
the all-sky High Time Resolution Universe (HTRU) surveys
(\citeauthor{2010MNRAS.409..619K}~2010, Barr et al. in prep) are fully
searched for binaries (\citeauthor{2011yera.confE..29N}~2011, Ng et
al. in prep), or until surveys to be performed with the next
generation of radio telescopes, e.g. LOFAR, FAST, and the SKA
(e.g. \citeauthor{2011A&A...530A..80S}~2011, van Leeuwen \& Stappers
2010\nocite{2010A&A...509A...7V}, Smits et al.,
2009a,b\nocite{2009A&A...505..919S}\nocite{2009A&A...493.1161S},
Cordes et al., 2004\nocite{2004NewAR..48.1413C}) are completed.  Some
of the other points, relevant to this analysis, are now discussed in
more detail.

\subsection{Additional acceleration search algorithm tests}
\label{s:add_algorithm_tests}
After the previous re-processing of the PMPS (\citeauthor{fau04}~2004,
\citeauthor{fsk+04}~2004) it was understood that much of the
sensitivity of the survey to relativistic binary pulsars was being
lost because of two principal effects: firstly, the incoherent
addition of spectra in the stack acceleration search and secondly, no
compensation for the effects of jerk caused by large changes in
orbital phase throughout the 35 minute observations. These problems
have been partially addressed in this work by implementing a coherent
acceleration search over a wide range of trial accelerations, and upon
independent halves of the original 35 minute integrations. To
establish if any systematics in the PMACCN algorithm were responsible
for the non-detection of any new relativistic binary pulsars we have
performed a number of additional tests, which are described in this
section. Because the primary goal of this work has been to find
relativistic binary systems like the Double Pulsar (and those more
extreme binary systems) we have chosen this system as our test case.

\begin{figure*}
\centering
\includegraphics[width=5.75cm]{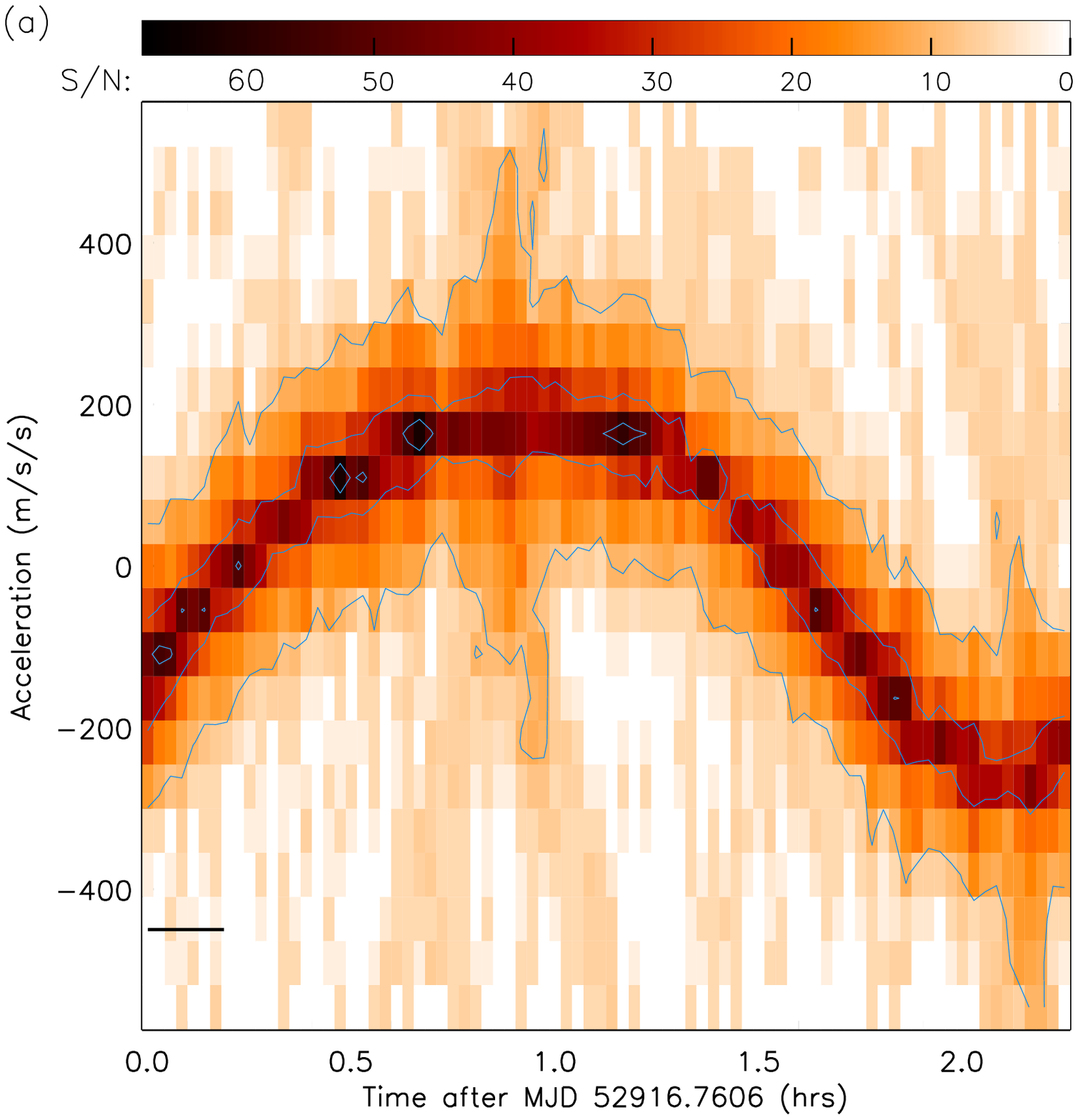}
\includegraphics[width=5.75cm]{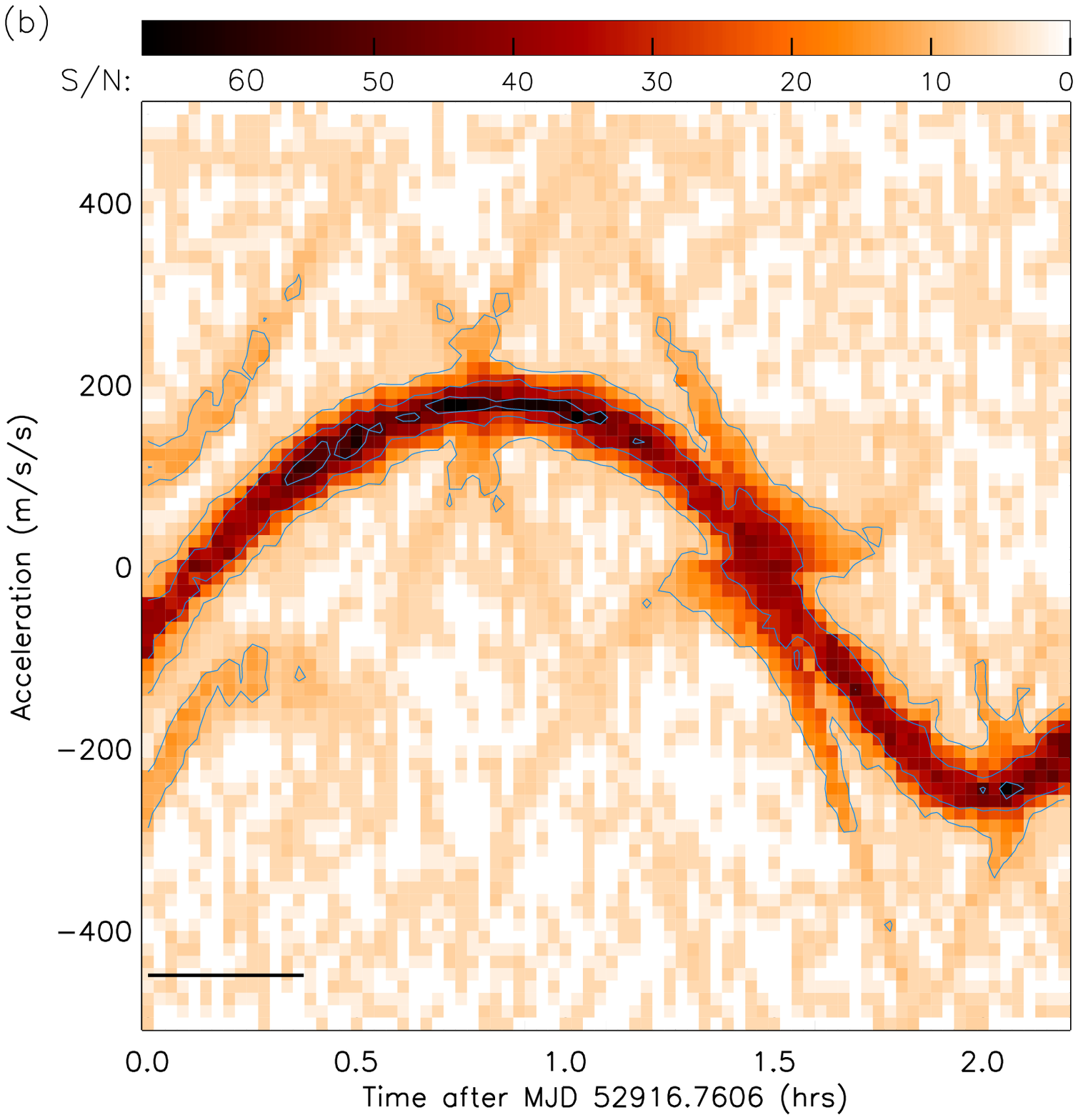}
\includegraphics[width=5.75cm]{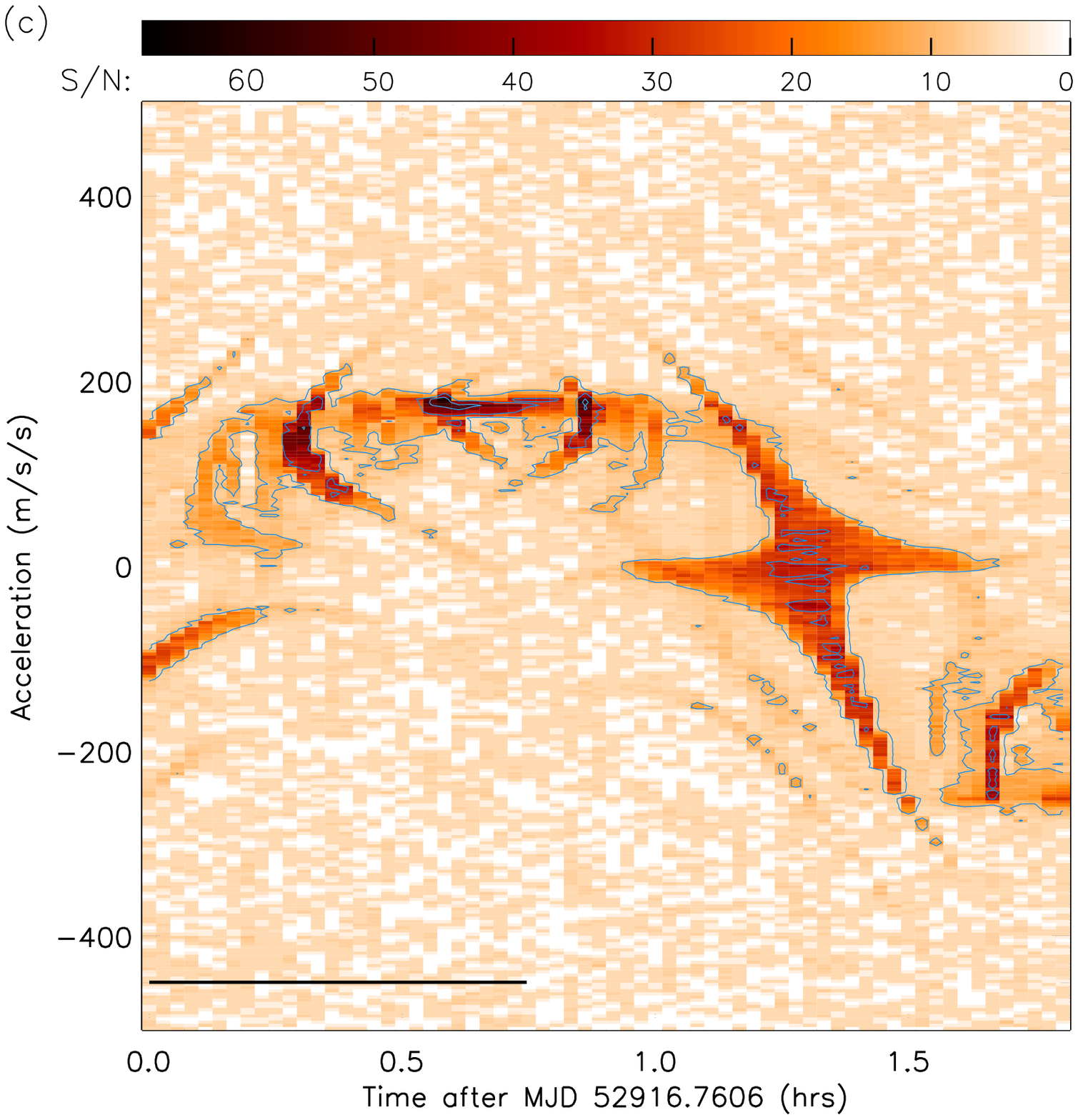}
\includegraphics[width=5.75cm]{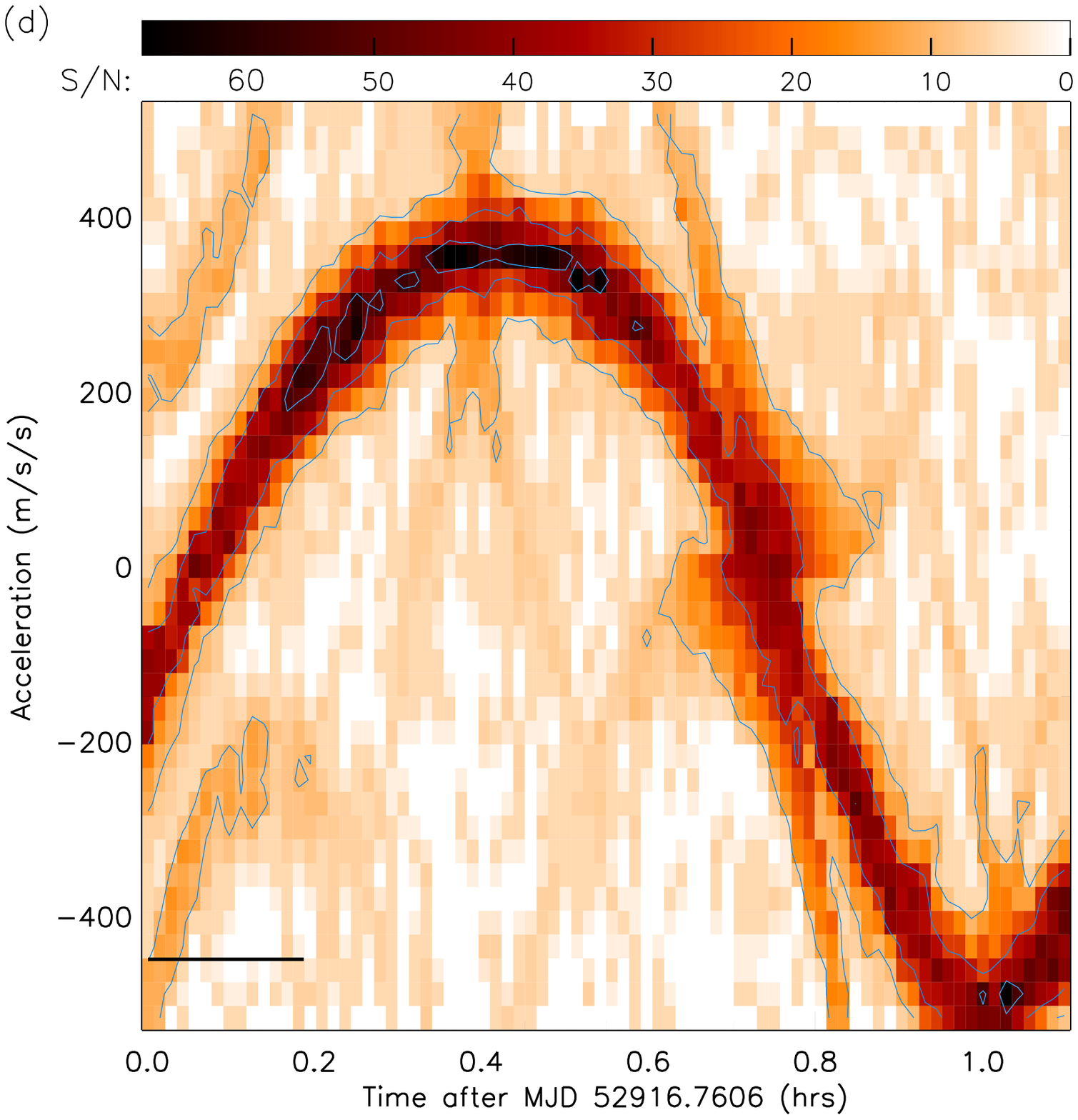}
\includegraphics[width=5.75cm]{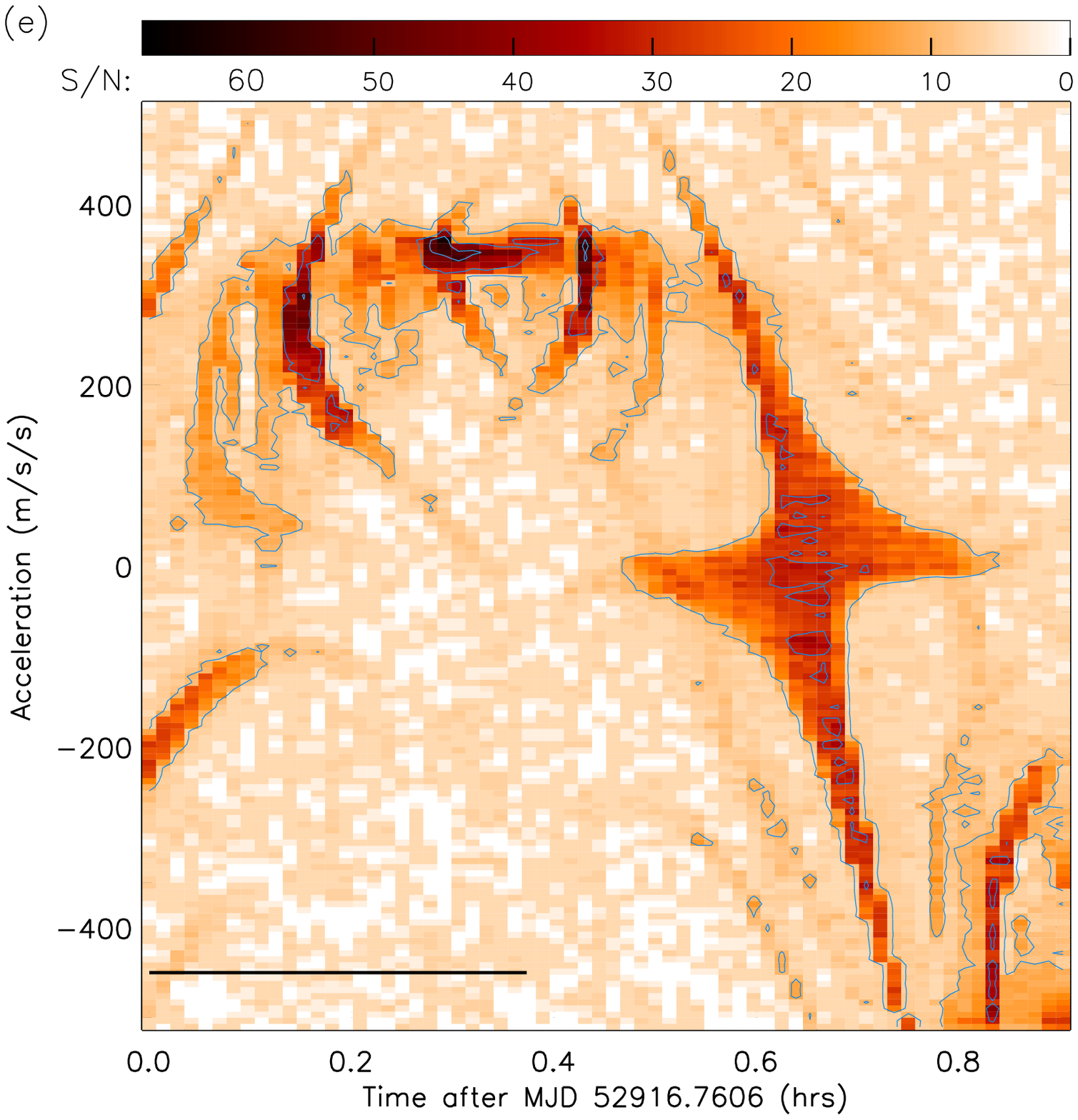}
\includegraphics[width=5.75cm]{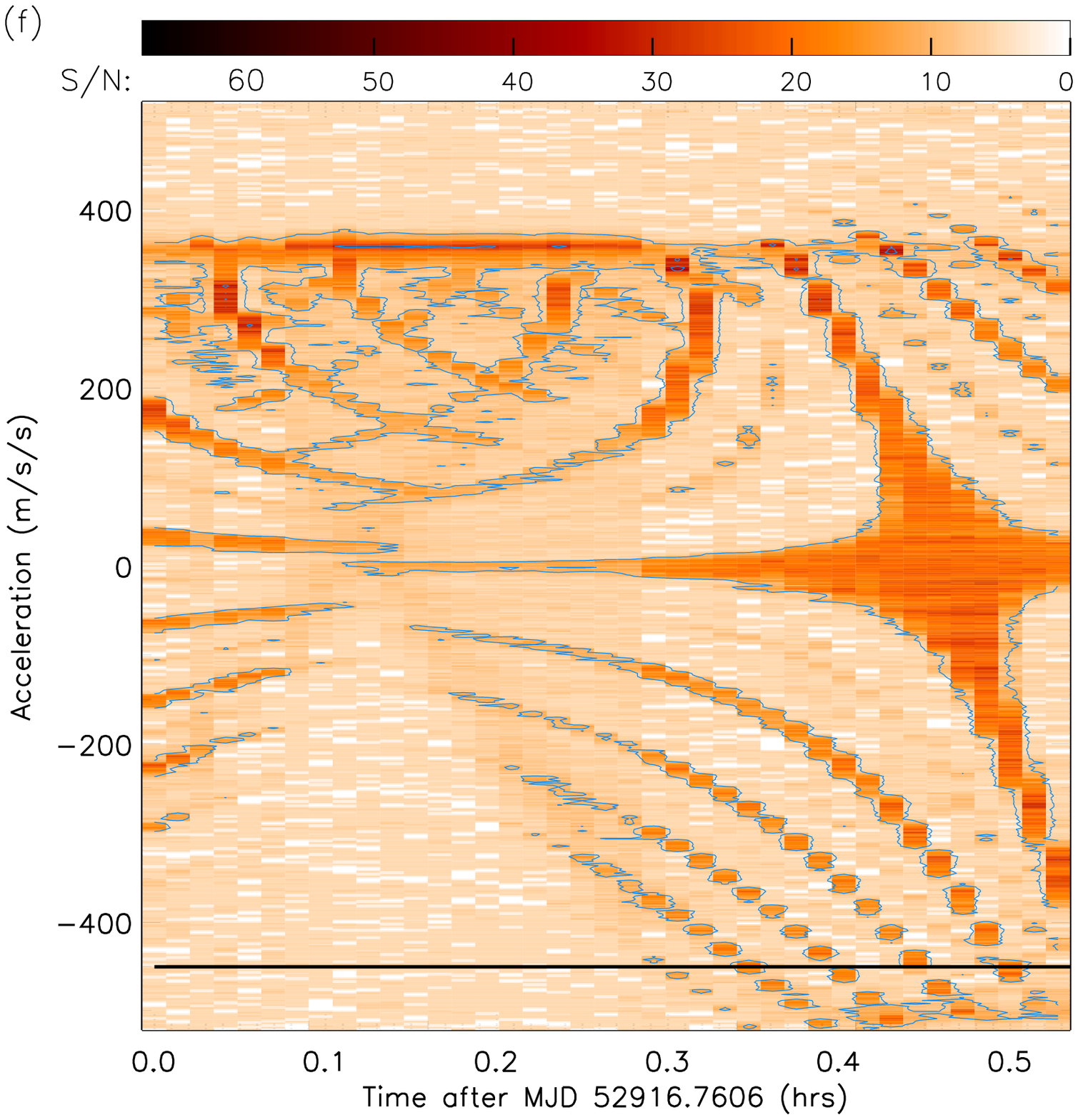}
\caption{The results from acceleration searches, using different
  integration lengths, across a 2.6 hour Parkes observation of the
  Double Pulsar system (panels (a), (b), and (c)) and the same data,
  but modified with a false sampling interval half that of the
  original to effect a $1.2$ hour binary (panels (d), (e), and
  (f)). Each panel shows the spectral S/N (of PSR J0737$-$3039A) as a function of
  trial acceleration for acceleration searches incremented in orbital
  phase by 100 seconds (panels (a), (b), and (c)) and 50 seconds
  (panels (d), (e), and (f)). The left-hand column of panels (panels
  (a) and (d)) show acceleration searches performed with an
  integration length of 671 seconds, the middle column (panels (b) and
  (e)) 1342 seconds, and the right-hand column (panels (c) and (f))
  2684 seconds. The black bar at the bottom of each panel indicates
  the length of the integration time used in each acceleration search.
  The maximum and average ${\rm S/N}_{\rm peak}$ are summarized in
  Table~A1 and are given here as follows. Panel (a):
  62.6, 43.7, panel (b): 66.9, 48.3, panel (c): 62.8, 32.4, panel (d):
  66.6, 48.3, panel (e), 62.8, 32.4, panel (f): 36.4, 29.1. Contours,
  plotted on top of the colour scale, mark spectral S/Ns of 10, 30,
  and 60.}
\label{f:search_tests}
\end{figure*}

\subsubsection{Integration length analysed and orbital phase effects}
\label{s:int_length_orb_phase}
Since the acceleration search technique can only correct for the
quadratic drift in pulse phase (and the corresponding linear drift in
pulse frequency) caused by a constant acceleration, sensitivity would
have been lost to any systems that showed changes in acceleration
during the PMPS observation. Without performing the higher order
corrections necessary to remove the effects of jerk, the only method
by which this can be mitigated in an acceleration search is to reduce
the integration length analysed (\citeauthor{jk91}~1991). In
  addition, the orbital phase at which the pulsar is observed will
  determine how well the constant acceleration assumption can model
  the data.  To study both the effects of the integration length
  analysed (to find if the 1050 s integration time used in this work
  was optimal for the detection of relativistic binary systems), and
  to probe the effects of orbital phase at which the pulsar is
  observed, the following tests have been performed: Acceleration
searches using different integration lengths, and separated by short
intervals in orbital phase, have been performed on archival
data, from the Parkes radio telescope, that covers one full orbit of
the Double Pulsar system.  Additionally, to create a more rigorous
test, the same data have been modified with a false sampling interval
half that of the original 80~$\mu$s. This has the effect of halving
both the pulse and orbital period, and doubling the apparent
line-of-sight acceleration to $\sim520\,{\rm m\;s^{-2}}$.

Three integration times, that are similar to small factors of the
original PMPS integration time, have been investigated: ``full-length"
2684 second segments, ``half-length" 1342 second segments, and
``quarter-length" 671 second segments\footnote{\footnotesize The
  integration times investigated do not match exact factors of the
  original PMPS observation time because of the use of a different
  data sampling interval, and a technical requirement in the tests for
  a total number of samples equal to a power of two.}. Appropriate
step sizes in acceleration trail for each integration length have been
computed using Equation~3. To investigate the effects of orbital phase
on the different length acceleration searches, the starting point of
each search has been incremented by 100 seconds for the original data,
and 50 seconds for the modified data across the entire orbit.

In Figure~\ref{f:search_tests} the results of these tests are
presented. Each panel shows the spectral S/N of PSR
  J0737$-$3039A as a function of trial acceleration for an
acceleration search started at a time after MJD 52916.76 given in
hours. Figure~\ref{f:search_tests} panels (a), (b) and (c) show the
results from the original, un-modified data.  Results from these tests
are also summarized in Table 6 in addition to results from tests
accounting for the effects of jerk
(Section~\ref{s:incl_jerk}). Acceleration searches of the quarter and
half-length segments (panels (a) and (b) respectively) are comparable
in their sensitivity over all orbital phases although there is a small
improvement in sensitivity for the half-length segments, which show
both a higher maximum and average peak spectral S/N (${\rm S/N}_{\rm
  peak}$) across the orbit\footnote{\footnotesize ${\rm S/N}_{\rm
    peak}$ refers to the best spectral S/N achieved in the
  acceleration search. The maximum ${\rm S/N}_{\rm peak}$ is the
  highest value found in the acceleration searches across the
  orbit. The average ${\rm S/N}_{\rm peak}$ is an average of all
  acceleration searches performed across the orbit.}. Importantly, the
average ${\rm S/N}_{\rm peak}$ in the half-length segments is larger
only by a factor of $\sim1.1$ compared to the quarter-length segments;
less than the expected increase of $\sqrt2$ given by the radiometer
equation. This can be explained by a reduction in sensitivity in the
half-length segments because of the increased effects of jerk. This
deleterious effect is confirmed by the fact that the maxima in ${\rm
  S/N}_{\rm peak}$ occur at integrations that cover orbital phases
where the acceleration is closest to constant, viz. at the peaks and
troughs. The corresponding minima in ${\rm S/N}_{\rm peak}$ occur at
orbital phases where the acceleration is changing most rapidly. In the
quarter-length analysis (panel (a)), where the acceleration can be
considered constant at almost all orbital phases, the maxima in ${\rm
  S/N}_{\rm peak}$ occur more evenly across the orbit.

As expected, the effectiveness of acceleration analyses using the
full-length segments (panel (c)), is strongly dependent upon orbital
phase.  Although a maximum ${\rm S/N}_{\rm peak}$, comparable to those
in the shorter analyses, can be achieved, the average ${\rm S/N}_{\rm
  peak}$ is significantly reduced by a factor of $\sim1.5$ with
respect to half-length analyses.

Figure~\ref{f:search_tests} panels (d), (e), and (f) show the results
from acceleration searches of the data modified with a false sampling
interval and using the same integration length segments as
previously. In this data, where the effective orbital period is now
$1.2$ hours, only acceleration searches that are performed on the
short quarter-length segments (panel (d)) have sensitivity across all
of the orbit. Both the values of the maximum and average ${\rm
  S/N}_{\rm peak}$ in this search (66.6 and 48.3 respectively) and the
occurrence of maxima in ${\rm S/N}_{\rm peak}$ at orbital phases where
the acceleration is changing slowly (peaks and troughs) make this the
direct analogue of the half-length analysis of the un-modified data
(panel (b)), but with an acceleration range expanded by a factor of
two. The same situation applies to panels (c) and (e). Therefore, an
analysis based on 336~s integrations on the modified data should
repeat the results of the 671~s analysis of the un-modified data
(panel (a)). Tests using full-length integrations on the modified data
produce the lowest values of maximum and average ${\rm S/N}_{\rm
  peak}$ from all the tests performed. Because of the relative
brightness of PSR J0737$-$3039A, detections can still be made at most orbital
phases, however compared to the optimum acceleration search of the
modified data (671~s segments) the average ${\rm S/N}_{\rm peak}$ is
reduced by 40 per cent.

From these tests it is clear that the effectiveness of acceleration
searches for the detection of `Double Pulsar like' systems in long
integration lengths ($T_{\rm obs}\gtrsim2000$ s) are highly dependent
upon the orbital phase at which the pulsar is observed. With the
integration length used in this work (1050 s) the sensitivity to these
systems lies somewhere in between the tests displayed in
Figure~\ref{f:search_tests} panels (a) and (b), suggesting this search
should be sensitive to such systems at most orbital phases. For
systems like the modified Double Pulsar system ($P_{\rm b} = 1.2~{\rm
  hrs}$, maximum acceleration $\sim520~{\rm m\;s^{-2}}$) the
possibility of detection in this work is strongly dependent on the
orbital phase at which the pulsar was observed; the search is most
effective when the pulsar is maximally accelerated.

\begin{figure}
\includegraphics[scale=0.3, angle=-90]{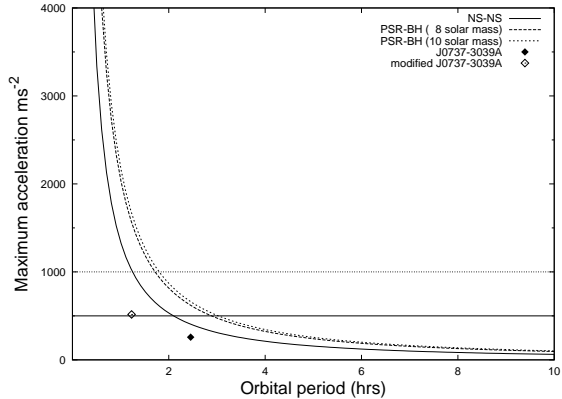}
\caption{Maximum line-of-sight orbital acceleration as a function of
  orbital period for circular edge on binary systems. The neutron
  stars have a mass of 1.3~$M_{\odot}$ and the two black hole
  companions have a mass of 8~$M_{\odot}$ and 10~$M_{\odot}$. The
  solid horizontal line shows the magnitude of the maximum constant
  acceleration that this search was sensitive to of $\pm 500\;{\rm
    m\;s}^{-2}$. The diamond points show the maximum acceleration of
  PSR J0737$-$3039A in the Double Pulsar system and the ``modified''
  Double Pulsar system described in Section~6.1.1.}
\label{f:accel_sens}
\end{figure}

\subsubsection{Acceleration search range}
\label{s:acc_search_range}
Neglecting the effects of jerk, and only considering the magnitude of
the orbital accelerations that might be present in even more compact
DNS and NS-BH systems, it can be shown that the search performed in
this work would not be capable of detecting these
systems. Figure~\ref{f:accel_sens} shows the maximum line-of-sight
acceleration as a function of orbital period, for circular edge on
binary systems, as given by Kepler's laws. In all curves the neutron
stars have a mass of 1.3~$M_{\odot}$ and the two black hole companions
have a mass of 8~$M_{\odot}$ and 10~$M_{\odot}$. The solid horizontal
line shows the magnitude of the maximum constant acceleration that
this search was sensitive to of $\pm 500\;{\rm m\;s}^{-2}$.  The
maximum accelerations present in the Double Pulsar system and in the
`modified' Double Pulsar system, described in Section~6.1.1, are
indicated by the diamond points.

Using the PMACCN algorithm the most compact circular NS-BH system that
might be detectable would in be in an orbit of about three hours.  To
have an improved chance of finding the most extreme NS-BH systems
($P_{\rm b}\gtrsim 2$ hr) the acceleration search range should reach
at least $\pm 1000\;{\rm m\;s}^{-2}$ (dotted horizontal
line). Performing an acceleration search over this range, and with a
computational cost no more than that used in the PMACCN algorithm,
could be achieved by segmenting the PMPS data into four 525 s segments
per beam. 

\begin{table*}
\label{t:accel_tests}
\caption{Summary table of results from acceleration and acceleration
  and jerk searches on the Double Pulsar system (J0737$-$3039A) and
  the ``modified'' Double Pulsar system (J0737$-$3039A$^{\bf
    *}$). ${\rm S/N}_{\rm peak}$ refers to the best spectral S/N
  achieved in the acceleration and acceleration and jerk search. The
  number of acceleration (or acceleration and jerk) trials at each DM
  searched, $n_{\rm trials}$/DM, is quoted for a downsampled sampling
  interval used in each test of 1280~$\mu$s for J0737$-$3039A and
  640~$\mu$s for J0737$-$3039A$^{\bf *}$.}  
\centering
\begin{small}
\begin{tabular}{rrccrcc}\\\hline\hline
\multicolumn{1}{c}{$T_{\rm obs}$ (s)} & \multicolumn{3}{c}{(acceleration searches)} & \multicolumn{3}{c}{(acceleration $+$ jerk searches)}\\
& $n_{\rm trials}$/DM & Maximum S/N$_{\rm peak}$ & Average S/N$_{\rm peak}$ & $n_{\rm trials}$/DM & Maximum S/N$_{\rm peak}$ & Average S/N$_{\rm peak}$ \\\hline
\multicolumn{1}{l}{J0737$-$3039A:} & & & & & &\\ 
 671 &  21 & 62.6 & 43.7 &   21 & 62.6 & 43.7\\
1342 &  75 & 66.9 & 48.3 &  375 & 66.9 & 51.9\\
2684 & 295 & 62.8 & 32.4 & 7965 & 84.2 & 53.8\\\hline
\multicolumn{1}{l}{J0737$-$3039A$^{\bf *}$:} & & & & & &\\
671 &   21 & 66.6 & 48.3 & - & - & - \\
1342 &  75 & 62.8 & 32.4 & - & - & - \\
2684 & 295 & 36.4 & 29.1 & - & - & - \\\hline
\end{tabular}
\end{small}
\end{table*}

\subsubsection{Including jerk}
\label{s:incl_jerk}
\begin{figure}
\begin{center}
\includegraphics[scale=0.32,angle=-90]{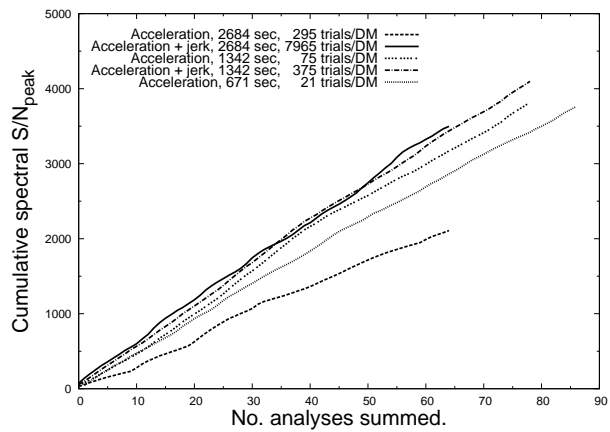}
\caption{The cumulative ${\rm S/N}_{\rm peak}$ for
  detections of PSR J0737$-$3039A from the acceleration and
  acceleration and jerk searches across one orbit of the Double Pulsar
  system.}
\label{f:jerk_effect}
\end{center}
\end{figure}

From an extension of the acceleration algorithm presented in Section~3
a step-size in jerk can be computed. The pulse broadening due to jerk,
$\tau_{\rm jerk}$, can be written,
\begin{equation}
\label{e:jbroad}
\tau_{\rm jerk}(t) = jt^3/6c\rm{,}
\end{equation}
where $j$ is the value of line-of-sight jerk, typically given in units
of ${\rm cm \, s}^{-3}$, $t$ is time and $c$ is the speed of light
(see for example Equation 4 in \citeauthor{jk91}~1991).  Following the
principles outlined in Section~3 the maximum acceptable value of
pulse broadening due to jerk at either end of the integration
should be $8 \tau_{\rm samp}$ for at least 50 per cent of the pulses
to be smeared by less than one time sample. A maximum pulse broadening
time of $8 \tau_{\rm samp}$ allows a step of $16 \tau_{\rm samp}$
between jerk trials. Setting Equation~\ref{e:jbroad} equal to this
value and once again letting $t=T_{\rm obs}/2$, the step-size in jerk,
$\delta j$ can be written as,
\begin{equation}
\label{e:jstep} 
\delta j = 768c\tau_{\rm samp}/T_{\rm obs}^3.
\end{equation}
To find the sensitivity to be gained by correcting for jerk we have
re-performed the tests described in
Section~\ref{s:int_length_orb_phase} on the un-modified Double Pulsar
data (Figure~\ref{f:search_tests} panels (a),(b), and (c)). Step sizes
have been computed with Equations~3 and~\ref{e:jstep} and a range in
jerk of $\pm 20\; {\rm cm\,s}^{-3}$ (c.f. jerk in Double Pulsar system
$\sim 18\;{\rm cm\,s}^{-3}$) has been searched. Again, the maximum and
average values of ${\rm S/N}_{\rm peak}$ across the orbit for the
different integration lengths are summarized in Table~6. Tests on the
`modified' Double Pulsar system have not been performed due to
limitations in available computing power. For the the shortest
quarter-length analysis (671 s) the results are the same as the
acceleration tests as no jerk trials are required within the range we
have chosen. For the half-length segments there is no change in the
maximum ${\rm S/N}_{\rm peak}$ achievable but the average ${\rm
  S/N}_{\rm peak}$ has increased by seven per cent. The biggest
improvement from the application of jerk corrections is in the
full-length (2684 s) segments where both the best maximum and average
${\rm S/N}_{\rm peak}$ from all of the searches are achieved. These
results can also be seen in Figure~\ref{f:jerk_effect} where the
cumulative value of ${\rm S/N}_{\rm peak}$ of PSR J0737$-$3039A from
the acceleration and acceleration and jerk searches across the Double
Pulsar orbit are plotted.

From Table~6 and Figure~\ref{f:jerk_effect} it can
also be seen that the improvement in average ${\rm S/N}_{\rm peak}$
across the orbit between acceleration and jerk searches of the
half-length and full-length integrations is small (factor of
$1.04$). Because there are a factor of $\sim21$ times more trials to
perform the acceleration and jerk analyses on the full-length segments
compared to the half-length segments, there appears to be large
computational expense with little gain in sensitivity. These lower
than expected gains in sensitivity by including jerk trials are most
likely due to poor approximation of the line-of-sight orbital motion
over the length of the integration. To combat these effects a third
derivative of velocity might need to be included.

\subsection{Pulsar candidate non-confirmations}
\label{s:nonconf}
\begin{figure*}
\begin{center}
\hrule
\vspace{0.2truecm}
\includegraphics[scale=0.66]{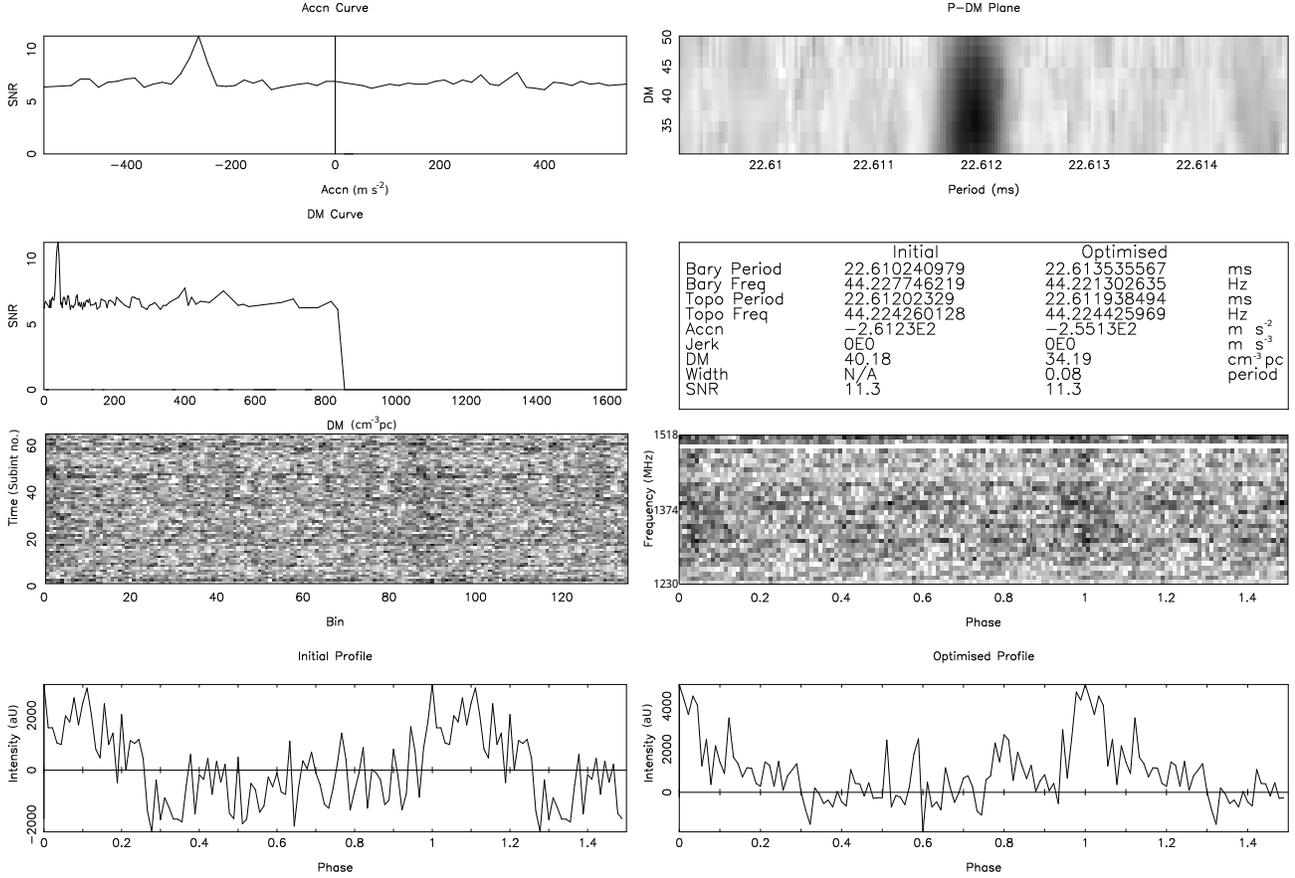}
\vspace{0.2truecm}
\hrule
\caption{The class three relativistic binary pulsar
  candidate, J1431$-$6042. Starting clockwise from the bottom left the
  candidate plot displays: the integrated pulse profile, folded at the
  initial spectral discovery period and DM; 64 temporal
  subintegrations of the observation, showing how the pulse varies
  with time; the spectral S/N as a function of a wide range of trail
  DMs; the spectral S/N as a function of trial acceleration (here the
  pulsar candidate has been detected with an acceleration of
  $-255\;{\rm ms}^{-2}$); the period-DM diagram, that shows how the
  S/N varies with small changes in the folding period and DM; the
  basic discovery parameters before and after time domain
  optimisation; stacked pulses across 32 frequency subbands, showing
  how the pulse varies with observing frequency, and finally the
  integrated pulse profile, folded at the optimum period and DM.}
\label{f:nonconf_eg}
\end{center}
\end{figure*}
Although no relativistic binary systems have been discovered the
acceleration search has produced a number of high-quality candidates
that have not been confirmed, despite multiple re-observations at
either the Parkes or Jodrell Bank
observatories. Figure~\ref{f:nonconf_eg} shows an example of one such
source, J1431$-$6042, a highly accelerated class three candidate. This
candidate has been followed up with a `grid' observation
(e.g. \citeauthor{mhl+02}~2002) and around four hours of
re-observations at Parkes. Because the orbital phase and hence the
orbital acceleration are likely to be different upon re-observation
all the data have been searched in acceleration on the TIER2 facility
with the PMACCN algorithm, and independently with the {\sc presto}
software suite\footnote{\footnotesize
  http://www.cv.nrao.edu/$\sim$sransom/presto/} over a wider range of
accelerations and including corrections for jerk. To date no
conclusive re-detection of this candidate has been made.

In total there are 12 other candidates for which a similar situation
applies. The majority come from our class two and class three
candidate lists. Typically these candidates appear accelerated, have a
S/N near our detection threshold (${\rm S/N}\sim 9$), and are often
only found in one of the half integrations. Many of these detections
might be attributed to strong `de-accelerated' RFI signals. Strong
periodic RFI can appear in the higher DM trials and sometimes even
swept in radio frequency mimicking a dispersed signal. In addition,
RFI signals are typically unstable causing them to drift through a
number of spectral bins in the fluctuation spectrum.

If any of these candidates are in fact true celestial sources the
systems may have undergone geodetic precession, where the spin axis of
the pulsar precesses with respect to the total angular momentum of the
system, which is dominated by the orbital angular momentum
\citep{mydr74}.  In a manner similar to that shown for PSR B1913$+$16
by \cite{kra98}, the effect may have shifted the orientation of
emission beam away from the Earth over the $\sim\,$10 years since the
original PMPS observation.  This is the current explanation for why
the relativistic binary pulsar, J1141$-$6545 \citep{klm+00a} was not
detected in an earlier 70 cm survey that covered the position
(\citeauthor{mks+10}~2010). We can estimate the magnitude of geodetic
precession for candidates that are close to the orbital period limit
of our search. Assuming a minimum detectable orbital period of $\sim
3$ hrs (using the PMACCN search algorithm) gives a precessional rate
of 3.5$^\circ$ yr$^{-1}$ for circular DNS systems (neutron star mass
of 1.3~$M_{\odot}$) and 14.7$^\circ$ yr$^{-1}$ for circular NS-BH
systems (black hole mass of 10.0~$M_{\odot}$). Although the alignment
between the spin axis of the pulsar and orbital angular momentum axis
is important in determining by how much the emission beam will move,
it is possible that relativistic systems detected in the PMPS survey
observations may have precessed out of the line-of-sight in the
intervening years.

At the time of writing, observations to detect un-confirmed candidates
from this search are still performed when telescope time becomes
available.

\section{Summary \& future work}
\label{s:sum_future}
A full re-analysis of the PMPS has been performed using coherent
acceleration searches on independent halves of the original 35 minute
integrations. 16 pulsars have been discovered, including a binary
millisecond pulsar, an intermittent pulsar, and a pulsar with an
interpulse. No new relativistic binary pulsars have been discovered.

Searches for relativistic binaries are already underway in the HTRU
pulsar surveys (\citeauthor{2011yera.confE..29N}~2011, Ng et al. in
prep). The low-latitude portion of the survey will cover a thin strip
($|b| < 3.5^{\circ}$) along the Galactic plane with integration times
of over 70 minutes. For these long integration times acceleration
searches will not be sensitive to the most relativistic binary
pulsars; higher order effects will need to be taken into account (see
Section~\ref{s:incl_jerk}). Unfortunately, it can be shown that
searches in these higher order effects are still computationally
prohibitive for long integrations. Assuming a sampling interval of
$128\;\mu \rm{s}$ and using step-sizes in acceleration and jerk given
by Equations~\ref{e:accstep} and \ref{e:jstep} respectively,
corrections for the effects of both acceleration and jerk in a `Double
Pulsar like' system would require of the order of $\sim10^7$ trials
per DM trial for a 70 minute observation (c.f. 59 acceleration trials
per DM as in this work). For the analysis of the low-latitude HTRU
data, either incoherent methods such as stack searches (Eatough et
al. in prep), or coherent acceleration searches of different length
but shorter segments of the original 70 minute observations should be
performed (Ng et al. in prep). 

Because of the limitations of the acceleration search performed in
this work, a full re-processing of the PMPS has been conducted
utilizing the Einstein@Home network\footnote{\footnotesize
  http://einstein.phys.uwm.edu/}. This search applies matched filters
to coherently de-modulate the data based on a number of Keplerian
orbital templates, which allows the full integration length to be
searched (Knispel et al. in prep).

The discovery and timing of highly relativistic binary pulsars is one
of the key science projects of the SKA
(\citeauthor{2004NewAR..48.1413C}~2004). As well as offering highly
rewarding discoveries, current searches for such systems, both in
archival data and in data from the latest pulsar surveys, will provide
considerable information on the binary parameter space still to be
searched in, and on the strategy of future pulsar surveys.

\section*{ACKNOWLEDGEMENTS}
This research was partly funded by grants from the Science \&
Technology Facilities Council. The Parkes radio telescope is part of
the Australia Telescope National Facility which is funded by the
Commonwealth of Australia for operation as a National Facility managed
by CSIRO. This work was based on observations with the 100-m telescope
of the MPIfR (Max-Planck-Institut f\"ur Radioastronomie) at
Effelsberg.  We would like to thank Xu D. D. for many useful
discussions about the ideas presented here. We thank C. Jordan and the
staff of the Parkes observatory for observational help at Jodrell Bank
and Parkes respectively. We also wish to thank E. F. Keane for
manuscript reading and A. Noutsos for useful discussions. The authors
kindly wish to thank A. Forti and The University of Manchester High
Energy Physics group for use of the TIER2 computing facility.

\bibliographystyle{mnras}

\end{document}